\documentclass[preprint,showpacs,preprintnumbers,amsmath,amssymb]{revtex4}

\usepackage{graphicx}
\usepackage{dcolumn}
\usepackage{bm}

\usepackage{feynmf}
\usepackage{slashed}

\usepackage[utf8]{inputenc}

\begin{document}

\title{NMSSM extended with vectorelike particles and the diphoton excess at the LHC}
\author{Yi-Lei Tang}
\thanks{tangyilei15@pku.edu.cn}
\affiliation{Center for High Energy Physics, Peking University, Beijing 100871, China}


\author{Shou-hua Zhu}
\thanks{shzhu@pku.edu.cn}
\affiliation{Institute of Theoretical Physics $\&$ State Key Laboratory of Nuclear Physics and Technology, Peking University, Beijing 100871, China}
\affiliation{Collaborative Innovation Center of Quantum Matter, Beijing 100871, China}
\affiliation{Center for High Energy Physics, Peking University, Beijing 100871, China}

\date{\today}

\date{\today}

\begin{abstract}

We investigate the parameter space of a model which extends next to minimal supersymmetric standard model (NMSSM) with the vectorlike (VL) particles \cite{MyPaper}. We find that the $10+\overline{10}$ model can explain the possible diphoton excess recently revealed by the ATLAS and CMS collaborations, although the predicted signal strength is a little smaller than the observed one.

\end{abstract}


\maketitle
\section{Introduction} \label{Introduction}

Recently, both the ATLAS and CMS collaborations probed the possible diphoton resonance near $m_{\gamma \gamma} = 750 \text{ GeV}$ \cite{ATLASDiPhoton, CMSDiPhoton}. Regardless of the insufficient significance of the data at present, theoretical efforts have been made \cite{DiPhoton1, DiPhoton2, DiPhoton3, DiPhoton4, DiPhoton5, DiPhoton6, DiPhoton7, DiPhoton8, DiPhoton9, DiPhoton10, DiPhoton11, DiPhoton12, DiPhoton13, DiPhoton14, DiPhoton15, DiPhoton16, DiPhoton17, DiPhoton18, DiPhoton19, DiPhoton20, DiPhoton21, DiPhoton22, DiPhoton23, DiPhoton24, DiPhoton25, DiPhoton26, DiPhoton27, DiPhoton28, DiPhoton29, DiPhoton30, DiPhoton31, DiPhoton32, DiPhoton33, DiPhoton34, DiPhoton35, DiPhoton36, DiPhoton37, DiPhoton38, DiPhoton39, DiPhoton40, DiPhoton41, DiPhoton42, DiPhoton43, DiPhoton44, DiPhoton45, DiPhoton46, DiPhoton47, DiPhoton48, DiPhoton49, DiPhoton50, DiPhoton51, DiPhoton52, DiPhoton53, DiPhoton54, DiPhoton55, DiPhoton56, DiPhoton57, DiPhoton58, DiPhoton59, DiPhoton60, DiPhoton61, DiPhoton62, DiPhoton63, DiPhoton64, DiPhoton65, DiPhoton66, DiPhoton67, DiPhoton68, DiPhoton69, DiPhoton70, DiPhoton71, DiPhoton72, DiPhoton73, DiPhoton74, DiPhoton75, DiPhoton76, DiPhotonAdded1, DiPhotonAdded2, DiPhotonAdded3, DiPhotonAdded4, DiPhotonAdded5, DiPhotonAdded6, DiPhotonAdded7, DiPhotonAdded8, DiPhotonAdded9, DiPhotonAdded10, DiPhotonAdded11, DiPhotonAdded12, DiPhotonAdded13, DiPhotonAdded14, DiPhotonAdded15, DiPhotonAdded16, DiPhotonAdded17, DiPhotonAdded18, DiPhotonAdded19, DiPhotonAdded20, DiPhotonAdded21, DiPhotonAdded22, DiPhotonAdded23, DiPhotonAdded24, DiPhotonAdded25, DiPhotonAdded26, DiPhotonAdded27, DiPhotonAdded28, ForceToAdd} and perhaps the most straightforward approach is to introduce a scalar field together with some additional vectorlike (VL) particles. Just similar to the production of the standard model (SM) like Higgs boson, the exotic scalar particle is produced through gluon-gluon fusion process induced by the vectorlike quark loops, and subsequently decays into gamma-gamma final states induced by the vectorlike particle loops. However, as for many specific models, it is fairly difficult to enhance the signal strength of the diphoton channel up to $\sim 10 \text{ fb}$ , which competes at least with the di-gluon, $Z \gamma$, and other tree-level $ZZ$, $W^+ W^-$, $h_{SM}h_{SM}$ decay channels. Sometimes other fermionic final state channels, e.g., $t \overline{t}$, $b \overline{b}$, may also dominate the total width. Symmetries are sometimes utilized in order to forbid some final states. For example, ignoring the CP-violation effects, if the exotic scalar particle $A$ is CP-odd, its decay to $h_{SM} h_{SM}$ is forbidden. Further more, if $A$ does not carry the $U(1)_Y \times SU(2)_L \times SU(3)_C$ quantum charges, its decays to $ZZ$, $W^+ W^-$ are eliminated at tree level as well. Additionally, decays into standard model fermions need to be further taken care of. However, in some specific models, there are quite a number of exotic fields, which need to be examined carefully.

Rather than building new models specifically for explaining the possible 750 GeV resonance, it is interesting to investigate through the existing models motivated by other problems. Supersymetric models (For a review, see \cite{MSSMInt}) can solve the hierarchy problem by adding each particle with a super-partner, which cancels the quadratic divergences in the Higgs self-energy diagrams. Unfortunately, within the framework of the minimal supersymetric standard model (MSSM), explaining the possible 750 GeV resonance is far from possible. However, in the Ref.~\cite{MyPaper}, one of the authors has proposed a supersymmetric model which combined the next to minimal supersymmetric standard model (NMSSM) (For reviews, see \cite{MSSMInt, NMSSMInt}) together with the VL particles (For examples, see \cite{VL1, VL2, VL3, VL4, VL5, VL6, VL7, VL8, VL9, VL10, VL11, VL12, VL13, VL14, VL15, VLAdded1, VLAdded2, VLAdded3, VLAdded4}). In this model, VL masses originate from the vacuum expectation value (VEV) of the singlet Higgs field, which offers a possible dynamical explanation to the origin of VL mass terms. Vectorlike particles contribute to the Higgs masses via the loop diagrams. On the other hand, extra Yukawa couplings adjust the trajectories of the renormalization group (RG) flow of the gauge coupling constants, helping the interactions to unify. In order not to bother the unification of the gauge coupling constants, which is an important feature of the supersymmetric models, complete $SU(5)$ multiplets are introduced. In the Ref.~\cite{MyPaper}, we discussed only the $5+\overline{5}$, $10+\overline{10}$, and $5+\overline{5}+10+\overline{10}$ cases. This model contains exactly an exotic scalar field together with some VL particles which couple strongly with the scalar field. However, in order not to counter Landau-pole below the unification scale, values of the Yukawa coupling constants are limited, resulting in a constraint of the $\gamma \gamma$ signal strength. In the following text of this paper, we go through the parameter space of this model and show that the $10+\overline{10}$ case provides a possible explanations to the diphoton excess, although the final signal strength seems to be a little bit smaller than the observed central value $\sim 10 \text{ fb}$. In fact one needs more data to determine the true value of signal strength. Historically, there are precedents that the best-fitted experimental diphoton signal strength exceed the predicted values \cite{Higgs1, Higgs2}.
\newpage	

\section{The Model and Some Conventions}

Since this model is based on the NMSSM, we write down the pure NMSSM part of the superpotential and the supersymmetry soft breaking terms
\begin{eqnarray}
W_{NMSSM} &=& \lambda H_u H_d S + \frac{\kappa}{3} S^3 + y_t Q_3 H_u U_3, \label{NMSSMSuperPotential}\\
V_{NMSSM}^{soft} &=& m_{Hu}^2 |\tilde{H}_u|^2 + m_{Hd}^2 |\tilde{H}_d|^2 + M_S^2 |\tilde{S}|^2 \nonumber \\
&+& (\lambda A_{\lambda} \tilde{H}_u \tilde{H}_d \tilde{S} + \frac{1}{3} \kappa A_{\kappa} \tilde{S}^3 + y_t A_{y_t} \tilde{Q}_3 \tilde{H}_u \tilde{U}_3+\text{h.c.}), \label{NMSSMSOftTerm}
\end{eqnarray}
where $H_{u,d}$ are the up and down type Higgs doublet. $S$ is the $U(1)_Y \times SU(2)_L \times SU(3)_C$ singlet Higgs superfield. Here we only show the couplings involving the top quark due to the relatively large coupling constant.

If we extend the NMSSM with the vectorlike particles, we need to introduce pairs of the vectorlike leptonic doublets $L$, $\overline{L}$, the vectorlike down-type quark singlets $D$, $\overline{D}$, the vectorlike quark doublets $Q$, $\overline{Q}$, the vectorlike up-type quark singlets $U$, $\overline{U}$, and the vectorlike charged lepton singlet $E$, $\overline{E}$. Their quantum numbers are listed in the Tab.~I of the Ref.~\cite{MyPaper}. In this paper, we only discuss the $5+\overline{5}$ model and the $10+\overline{10}$ model. Their superpotentials and supersymmetry soft breaking terms are listed below,
\begin{eqnarray}
W_{5+\overline{5}} &=& \lambda_D \bar{D} D S + \lambda_L \bar{L} L S. \label{5p5W} \\
V^{soft}_{5+\overline{5}} &=& m_D^2 (\tilde{D} \tilde{D}^{\dagger} + \tilde{\bar{D}} \tilde{\bar{D}}^{\dagger}) + m_L^2 (\tilde{L} \tilde{L}^{\dagger}+\tilde{\bar{L}} \tilde{\bar{L}}^{\dagger}) + (\lambda_D A_{\lambda_D} \tilde{\bar{D}} \tilde{D} \tilde{S} + \lambda_L A_{\lambda_L} \tilde{\bar{L}} \tilde{L} S + \text{h.c.}); \label{5p5V}
\end{eqnarray}
and
\begin{eqnarray}
W_{10+\overline{10}} &=& \lambda_Q \bar{Q} Q S + \lambda_U \bar{U} U S + \lambda_E \bar{E} E S + y_U Q H_u U + y_{\bar{U}} \bar{Q} H_d \bar{U}. \label{10p10W} \\
V_{10+\overline{10}}^{soft}&=& m_Q^2 ( Q Q^{\dagger} + U U^{\dagger} ) + m_E^2 E E^{\dagger}
+ ( A_{\lambda_Q} \lambda_Q \tilde{\bar{Q}} \tilde{Q} \tilde{S} + A_{\lambda_U} \lambda_U \tilde{\bar{U}} \tilde{U} \tilde{S} \nonumber \\
&+& A_{\lambda_E} \lambda_E \tilde{\bar{E}} \tilde{E} \tilde{S} + A_{y_U} y_U \tilde{Q} \tilde{H_u} \tilde{U} + A_{y_{\bar{U}}} y_{\bar{U}} \tilde{\bar{Q}} \tilde{H_d} \tilde{\bar{U}} + \text{h.c.}). \label{10p10V}
\end{eqnarray}

After the Higgs fields acquire VEVs,
\begin{eqnarray}
H_u^0 &=& v_u + \frac{H_{uR} + i H_{uI}}{\sqrt{2}} \nonumber \\
H_d^0 &=& v_d + \frac{H_{dR} + i H_{dI}}{\sqrt{2}} \\
S &=& v_s + \frac{S_R + i S_I }{\sqrt{2}}, \nonumber
\end{eqnarray}
we acquire three CP-even Higgs fields
\begin{eqnarray}
h_i = S_{i1} H_{uR} + S_{i2} H_{dR} + S_{i3} S_R,
\end{eqnarray}
together with the CP-odd Higgs fields
\begin{eqnarray}
A=\cos{\beta} H_{uI} + \sin{\beta} H_{dI} \nonumber \\
G=-\sin{\beta} H_{uI} + \cos{\beta} H_{dI}, \label{ImaginaryPartMixing1}
\end{eqnarray}
and
\begin{eqnarray}
a_1 = P_{11} A + P_{12} S_I \nonumber \\
a_2 = P_{21} A + P_{22} S_I, \label{ImaginaryPartMixing2}
\end{eqnarray}
where $S_{ij}$, $P_{ij}$ are the mixing matrix elements, $\tan{\beta} = \frac{v_u}{v_d}$, and $G$ is the goldstone state to be rotated away.

\section{the LHC diphoton excess in the model}

In this model, it is mainly the extra quarks which contribute to the production and the decay of the 750 GeV scalar particle. The unwanted decay modes of this scalar particle should be avoided. If one of the CP-even Higgs field, say $h_2$ is the $750 \text{ GeV}$ resonance, it is difficult to avoid large branching ratio of $h_2 \rightarrow h_{\text{SM}} + h_{\text{SM}}$, thus the diphoton rate is severely suppressed. As a result, we need to choose between $a_1$ and $a_2$. Without loss of generality, we adopt the convention that $P_{12} > P_{11}$. From (\ref{NMSSMSuperPotential}) and (\ref{ImaginaryPartMixing1}) we know that the $A$ couples with the top quark, and in fact, it also couples with other SM fermions, then we need to choose the singlet-like CP-odd Higgs boson in order to eliminate large branching ratios to the SM fermions. That is to say, without loss of generality, we can choose $a_1$ when $|P_{12}|$ approaches $1$.

Unfortunately, the CP-odd scalar particles does not couple with the squarks, which lowers the signal strength. After integrating out all the particles in the loop, we acquire the effective operators
\begin{eqnarray}
\mathcal{L}_{\text{eff}} \supset \frac{e^2}{2 \Lambda_\gamma} a_1 F_{\mu \nu} \tilde{F}^{\mu \nu} + \frac{g_3^2}{2 \Lambda_g} a_1 G_{\mu \nu} \tilde{G}^{\mu \nu},
\end{eqnarray}
where $e$ and $g_3$ are the electro-magnetic and the QCD coupling constants. $\tilde{F},\tilde{G}_{\mu \nu} = \frac{1}{2} \epsilon_{\mu \nu \lambda \rho} F,G^{\lambda \rho}$. To calculate the $\Lambda_g$ and the $\Lambda_\gamma$, we use the following formulae \cite{HEFT1, HEFT2},
\begin{eqnarray}
\frac{1}{\Lambda_g} &=& \frac{1}{4 \pi^2 M_{a_1}} \sum_i \frac{1}{2} \frac{1}{\sqrt{\tau_i}} y_{fi} \arcsin^2 (\sqrt{\tau_i}), \nonumber \\
\frac{1}{\Lambda_\gamma} &=& \frac{1}{4 \pi^2 M_{a_1}} \sum_i N_{ci} Q_{fi}^2 \frac{1}{\sqrt{\tau_i}} y_{fi} \arcsin^2 (\sqrt{\tau_i}),
\end{eqnarray}
where $N_{ci}$ equals 1 or 3 for the $SU(3)_c$ singlet or triplet of the Dirac particle $i$, $Q_{fi}$ is the charge number, and $\tau_i = \frac{M_{a_1}^2}{4 M_{i}^2}$ where $M_{i}$ is the mass of the Dirac particle $i$. The branching widths of the $a_1 \rightarrow gg$ and the $a_1 \rightarrow \gamma \gamma$ are given by
\begin{eqnarray}
\Gamma(a_1 \rightarrow g g) &=& 8 \pi \alpha_3^2 \frac{M_{a_1}^3}{\Lambda_g^2}, \nonumber \\
\Gamma(a_1 \rightarrow \gamma \gamma) &=& \pi \alpha^2 \frac{M_{a_1}^3}{\Lambda_g^2},
\end{eqnarray}
where $\alpha_3 = \frac{g_3^2}{4 \pi}$, and $\alpha = \frac{e^2}{4 \pi}$ is the fine-structure constant.

\subsection{The $5+\overline{5}$ Model}

For the $5+\overline{5}$ model, the loop diagrams involve $H_{u,d}$, $L$, $\overline{L}$, $D$, and $\overline{D}$. In order to avoid the Landau pole before the grand unification scale, the value of the Yukawa coupling constants are limited. We adopt the bench mark point
\begin{eqnarray}
\lambda_D(Q_{\text{GUT}})=3,~~~\lambda_L(Q_{\text{GUT}})=1.7,~~~\lambda(Q_{\text{GUT}})=3.0, ~~~Q_{\text{GUT}} = 1.8 \times 10^{16},
\end{eqnarray}
as the boundary condition, and do the renormalization group (RG) running down to the scale $Q=1 \text{ TeV}$ with the formula listed in the appendix of the Ref.~\cite{MyPaper}, then
\begin{eqnarray}
\lambda_D(1 \text{ TeV}) = 0.91,~~~\lambda_L (1 \text{ TeV}) = 0.44,~~~\lambda(1 \text{ TeV}) = 0.59.
\end{eqnarray}
We calculate the cross sections and the decay widths by the MadGraph5\_aMC@NLO v2.3.3. We adopt $v_s = 0.9 \text{ TeV}$ during the numerical calculation, and then we obtain $\sigma_{p p \rightarrow a_1} = 17.7 \text{ fb}$, $\Gamma(a_1 \rightarrow g g) = 2.47 \times 10^{-3} \text{ GeV}$, and $\Gamma(a_1 \rightarrow \gamma \gamma) = 8.84 \times 10^{-5} \text{ GeV}$. Even if there is no other decay modes, the signal strength $\sigma_{p p \rightarrow a_1} \times \text{Br}(a_1 \rightarrow \gamma \gamma) = 0.63 \text{ fb}$, which is far from explaining the observed excess.

One might think about adding several more copies of the $5+\overline{5}$ multiplets. In order not to encounter the Landau-pole until the gauge coupling constants unify, only limited number of copies can be added. From the Ref.~\cite{GMSB_GUT}, we learn that
\begin{eqnarray}
N_{5+\overline{5}} \lesssim 4.
\end{eqnarray}
For example, naively speaking, three copies of $D+\overline{D}$ might enhance the signal strength by a factor of about 9. However, due to the interaction terms between different couplings in the $\beta$-function, generally the Yukawa coupling constants near the TeV scale are further lowered, and what is worse, the $\lambda_L$ usually becomes so small that it is difficult to keep the masses of the vectorlike leptons above $\sim (750/2) \text{ GeV}$, which will open the $a_1 \rightarrow \overline{L} L$ channel and severely suppresses the branching ratio of the $a_1 \rightarrow \gamma \gamma$ channel.

\subsection{The $10+\overline{10}$ Model} \label{1010barSituation}

For the $10+\overline{10}$ model, the loop diagrams involve $Q$, $\overline{Q}$, $U$, $\overline{U}$, $E$, $\overline{E}$, and $H_{u,d}$. We adopt the boundary condition
\begin{eqnarray}
& & \lambda_Q(Q_{\text{GUT}})=0.5, ~~~\lambda_U(Q_{\text{GUT}})=0.52, ~~~\lambda_E(Q_{\text{GUT}})=1.1, ~~~\lambda(Q_{\text{GUT}})=1.6, \nonumber \\
& & \kappa (Q_{\text{GUT}})=3.5, ~~~Q_{\text{GUT}} = 3.6 \times 10^{16}.
\end{eqnarray}
as our bench mark point, then at $Q=1 \text{ TeV}$,
\begin{eqnarray}
& & \lambda_Q(1 \text{TeV})=0.73, ~~~\lambda_U(1 \text{TeV})=0.58, ~~~\lambda_E(1 \text{TeV})=0.46, ~~~\lambda(1 \text{TeV})=0.56, \nonumber \\
& & \kappa (Q_{1 \text{TeV}}) = 0.30 \label{Runned1010bar}
\end{eqnarray}
Again, we adopt $v_s = 0.9 \text{ TeV}$, then $\sigma_{p p \rightarrow a_1} = 186 \text{ fb}$, $\Gamma(a_1 \rightarrow g g) = 0.0261 \text{ GeV}$, and $\Gamma(a_1 \rightarrow \gamma \gamma ) = 0.000327 \text{ GeV}$. If there is no other decay mode, the signal strength $\sigma_{p p \rightarrow a_1} \times \text{Br}(a_1 \rightarrow \gamma \gamma) =  2.31 \text{ fb}$.


Compared with the $5+\overline{5}$ situation, there are more charged heavy quarks running in the loops, resulting in an enhancement of the production rate. The vectorlike $D$ particles take the charge of $\frac{1}{3}$, but in the $10+\overline{10}$, there are other two $\frac{2}{3}$-charged heavy quarks, which also result in the enhancement of the branching ratio of the $\gamma \gamma$ decay rate. However, due to the too much influence on the trajectories of the running gauge coupling constants, there leave us no much room to add copies of other $SU(5)$ multiplets to further enhance the signal strength.

We should also note that if $v_s$ is within our typical range of 700-900 GeV, the masses of the exotic quarks we introduce should lie roughly $\gtrsim 400$ GeV. According to the Ref.~\cite{PDG}, the current lower bounds on the exotic $t^{\prime}$, $b^{\prime}$ quarks can reach about 700 GeV. However, these constraints are all based on the assumption that the exotic quarks only mix with and then decay into the third generation SM quarks. If we only let the exotic quarks mix with the first two generation SM quarks, the lower bounds can be relaxed into $\sim 400$ GeV, which is compatible with our needed range. As for the charged exotic leptons, we can easily see that the bounds listed in the Ref.~\cite{PDG} are far below our needed range.

\section{The NMSSM tolerance}

More realistically, in the NMSSM, $a_1$ decays to other particles.  As has been mentioned, $a_1 \rightarrow h_{\text{SM}} h_{\text{SM}}$ is forbidden due to the conservation of the CP charge.  $a_1 \rightarrow a_j h_{1,2}$ is also forbidden kinematically if $m_{a_1} < m_{a_2}$. $a_1$ cannot decay into vectorlike particles and the Higgsinos because $2 m_{Q,U,D,\cdots} > m_{a_1}$. However, in order to eliminate $a_1 \rightarrow W^+ W^-$, $a_1 \rightarrow Z Z$, $a_1 \rightarrow t \overline{t}$, $a_1 \rightarrow b \overline{b}$, $\dots$, we can only hope $|P_{11}-1| \ll 1$ so that these channels are suppressed by the small mixings between the CP-even singlet-like Higgs and the doublet-like Higgs. We define the pure NMSSM width as $\Gamma_{a_1,\text{NMSSM}}$, which only sums over all the possible decay channels of $a_1$ calculated without the effects of the vectorlike sectors. If $\Gamma_{a_1,\text{NMSSM}} \gg \Gamma(a_1 \rightarrow g g)$, the signal strength will be highly suppressed, which of course should not be the case. In order to have a look at the $\Gamma_{a_1,\text{NMSSM}}$, we scan the NMSSM parameter space by the NMSSMTools 4.8.2 \cite{NMSSMTools1, NMSSMTools2, NMSSMTools3} within this area
\begin{eqnarray}
& & 400 \text{ GeV} < M_{1,2} < 800 \text{ GeV},~~~M_3 = 3 M_{1,2},~~~2<\tan \beta<15,~~~0.5 < \lambda < 0.65, \nonumber \\
& & 0.05 < \kappa < 0.32,~~~400 \text{ GeV} < \mu_{\text{eff}} = \lambda v_s < 800 \text{ GeV},~~~400 \text{ GeV} < M_A < 1000 \text{ GeV}, \nonumber \\
& &700 \text{ GeV} < M_P < 800 \text{ GeV}, ~~~\Omega_{\text{DM}} < 0.131,
\end{eqnarray}
where $M_{1,2,3}$ are the soft masses of the gauginos, $M_A$ is the diagonal doublet CP-odd mass matrix element, and $M_P$ is the diagonal singlet CP-odd mass matrix element. In order to forbid $a_1 \rightarrow 2h_X$, or $a_1 \rightarrow \text{doublet-like neutralinos}$ we set the lower limit of these mass parameters as 400 GeV, which is near $\frac{750 \text{ GeV}}{2}$. $a_1 \rightarrow \text{doublet-like neutralino+singlet-like neutralino}$ can be suppressed if the singlet-doublet neutralino mixings are small. We also set the relic density of the lightest neutralino $\Omega_{\text{DM}} < 0.131$ because the dark matter might not be composed of only one component. 

Vectorlike particles may modify the masses of the Higgs bosons. In this paper, we do not consider the $y_U$ and $y_{\overline{U}}$ and set them as zero for simplicity. Thus, only the mass of the singlet-like Higgs boson receives some loop-corrections. As for the $5+\overline{5}$ model, these corrections to the mass of the CP-even singlet-like Higgs is given by \cite{MyPaper},
\begin{eqnarray}
\Delta m_S^2 &=& \lambda_L^2 \frac{3 A_{\lambda_L} v_s^2 \kappa + 4 v_s^3 \kappa^2 + A_{\lambda_L} v^2 \sin{\beta} \cos{\beta} \lambda}{8 \pi^2 v_s} \ln{\frac{m_L^2}{Q^2}} \nonumber \\
&+& 3 \lambda_D^2 \frac{3 A_{\lambda_D} v_s^2 \kappa + 4 v_s^3 \kappa^2 + A_{\lambda_D} v^2 \sin{\beta} \cos{\beta} \lambda}{16 \pi^2 v_s} \ln{\frac{m_D^2}{Q^2}} \nonumber \\
&+& \frac{\lambda_L^4}{48 \pi^2 m_L^4}(-2 A_{\lambda_L}^4 v_s^2 + 24 A_{\lambda_L}^2 m_L^2 v_s^2 - 15 A_{\lambda_L}^3 vs^3 \kappa + 90 A_{\lambda_L} m_L^2 v_s^3 \kappa - 
 36 A_{\lambda_L}^2 v_s^4 \kappa^2 \nonumber \\
&+& 72 m_L^2 v_s^5 \kappa^2 - 35 A_{\lambda_L} v_s^5 \kappa^3 - 12 v_s^6 \kappa^4 + 3 A_{\lambda_L}^3 v_d v_s v_u \lambda - 18 A_{\lambda_L} m_L^2 v_d v_s v_u \lambda \nonumber \\
&+& 24 A_{\lambda_L}^2 v_d v_s^2 v_u \kappa \lambda - 48 m_L^2 v_d v_s^2 v_u \kappa \lambda + 45 A_{\lambda_L} v_d v_s^3 v_u \kappa^2 \lambda + 24 v_d v_s^4 v_u \kappa^3 \lambda \nonumber \\
&-& 9 A_{\lambda_L} v_d^2 v_s v_u^2 \kappa \lambda^2 - 12 v_d^2 v_s^2 v_u^2 \kappa^2 \lambda^2 - A_{\lambda_L} v_d^3 v_u^3 \lambda^3 + 24 m_L^4 v_s^2 \ln{\frac{m_L^2}{\lambda^2 v_s^2}}) + \label{Singlet_Mass} \\
&+& \frac{\lambda_D^4}{32 \pi^2 m_D^4}(-2 A_{\lambda_D}^4 v_s^2 + 24 A_{\lambda_D}^2 m_D^2 v_s^2 - 15 A_{\lambda_D}^3 vs^3 \kappa + 90 A_{\lambda_D} m_D^2 v_s^3 \kappa - 
 36 A_{\lambda_D}^2 v_s^4 \kappa^2 \nonumber \\
&+& 72 m_D^2 v_s^5 \kappa^2 - 35 A_{\lambda_D} v_s^5 \kappa^3 - 12 v_s^6 \kappa^4 + 3 A_{\lambda_D}^3 v_d v_s v_u \lambda - 18 A_{\lambda_D} m_D^2 v_d v_s v_u \lambda \nonumber \\
&+& 24 A_{\lambda_D}^2 v_d v_s^2 v_u \kappa \lambda - 48 m_D^2 v_d v_s^2 v_u \kappa \lambda + 45 A_{\lambda_D} v_d v_s^3 v_u \kappa^2 \lambda + 24 v_d v_s^4 v_u \kappa^3 \lambda \nonumber \\
&-& 9 A_{\lambda_D} v_d^2 v_s v_u^2 \kappa \lambda^2 - 12 v_d^2 v_s^2 v_u^2 \kappa^2 \lambda^2 - A_{\lambda_D} v_d^3 v_u^3 \lambda^3 + 24 m_D^4 v_s^2 \ln{\frac{m_D^2}{\lambda^2 v_s^2}}). \nonumber
\end{eqnarray}
The corrections to the CP-odd singlet-like Higgs boson are calculated as,
\begin{eqnarray}
\Delta m_{A}^2 &=& \frac{1}{4 \pi^2} \left[ \frac{3}{8} \frac{I_{AD}}{I_D} (I_D^- - I_D^+) + \frac{1}{4} \frac{I_{AL}}{I_L} (I_L^- - I_L^+) - \frac{3}{4} \lambda_D^2 (I_D^- + I_D^+) + \frac{3}{2} v_s^2 \lambda_D^4 \right. \nonumber \\
& & \left.  - \frac{1}{2} \lambda_L^2 (I_L^- + I_L^+) + v_s^2 \lambda_L^4 + \frac{3}{8} \frac{I_{AD}}{I_D} ( I_D^+ \ln \left( \frac{I_D^+}{Q^2} \right) - I_D^- \ln \left( \frac{I_D^-}{Q^2} \right) ) \right. \nonumber \\
& & \left. + \frac{3}{4} \lambda_D^2 (I_D^- \ln \left( \frac{I_D^-}{Q^2} \right) + I_D^+ \ln \left( \frac{I_D^+}{Q^2} \right) ) + \frac{1}{4} \frac{I_{AL}}{I_L} (I_L^+ \ln \left( \frac{I_L^+}{Q^2} \right) - I_L^- \ln \left( \frac{I_L^-}{Q^2} \right) ) \right. \nonumber \\
& & \left. \frac{1}{2} \lambda_L^2 ( I_L^- \ln \left( \frac{I_L^-}{Q^2} \right) + I_L^+ \ln \left( \frac{I_L^+}{Q^2} \right) ) - \frac{3}{2} v_s^2 \lambda_D^4 \ln \left( \frac{v_s^2 \lambda_D^2}{Q^2} \right) - v_s^2 \lambda_L^4 \ln \left( \frac{v_s^2 \lambda_L^2}{Q^2} \right) \right],
\end{eqnarray}
where
\begin{eqnarray}
I_D &=& | \lambda_D (A_{\lambda_D} v_s + v_s^2 \kappa - v_u v_d \lambda) |, \nonumber \\
I_L &=& | \lambda_L (A_{\lambda_L} v_s + v_s^2 \kappa - v_u v_d \lambda) |, \nonumber \\
I_D^+ &=& I_D + m_D^2 + v_s^2 \lambda_D^2, \nonumber \\
I_D^- &=& -I_D + m_D^2 + v_s^2 \lambda_D^2, \nonumber \\
I_L^+ &=& I_L + m_L^2 + v_s^2 \lambda_L^2, \nonumber \\
I_L^- &=& -I_L + m_L^2 + v_s^2 \lambda_L^2 \nonumber \\
I_{AD} &=& A_{\lambda_D}^2 \lambda_D^2 + 2 A_{\lambda_D} v_s \kappa \lambda_D^2 + 2 v_s^2 \kappa^2 \lambda_D^2 + 2 v_d v_u \kappa \lambda \lambda_D^2 \nonumber \\
I_{AL} &=& A_{\lambda_L}^2 \lambda_L^2 + 2 A_{\lambda_L} v_s \kappa \lambda_L^2 + 2 v_s^2 \kappa^2 \lambda_L^2 + 2 v_d v_u \kappa \lambda \lambda_L^2.
\end{eqnarray}
The related formula in the $10+\overline{10}$ case is too complicated to be listed in this paper. However, if we ignore all the vectorlike A-terms $A_{\lambda_D}$, $A_{\lambda_L}$, $A_{\lambda_Q}$, $A_{\lambda_U}$, $A_{\lambda_E}$, the typical corrections to the CP-odd singlet-like Higgs boson is about a few percent, which can be ignored under the current experimental data. Therefore, during the scanning process, we ignore these corrections.

\begin{figure}
\includegraphics[width=3in]{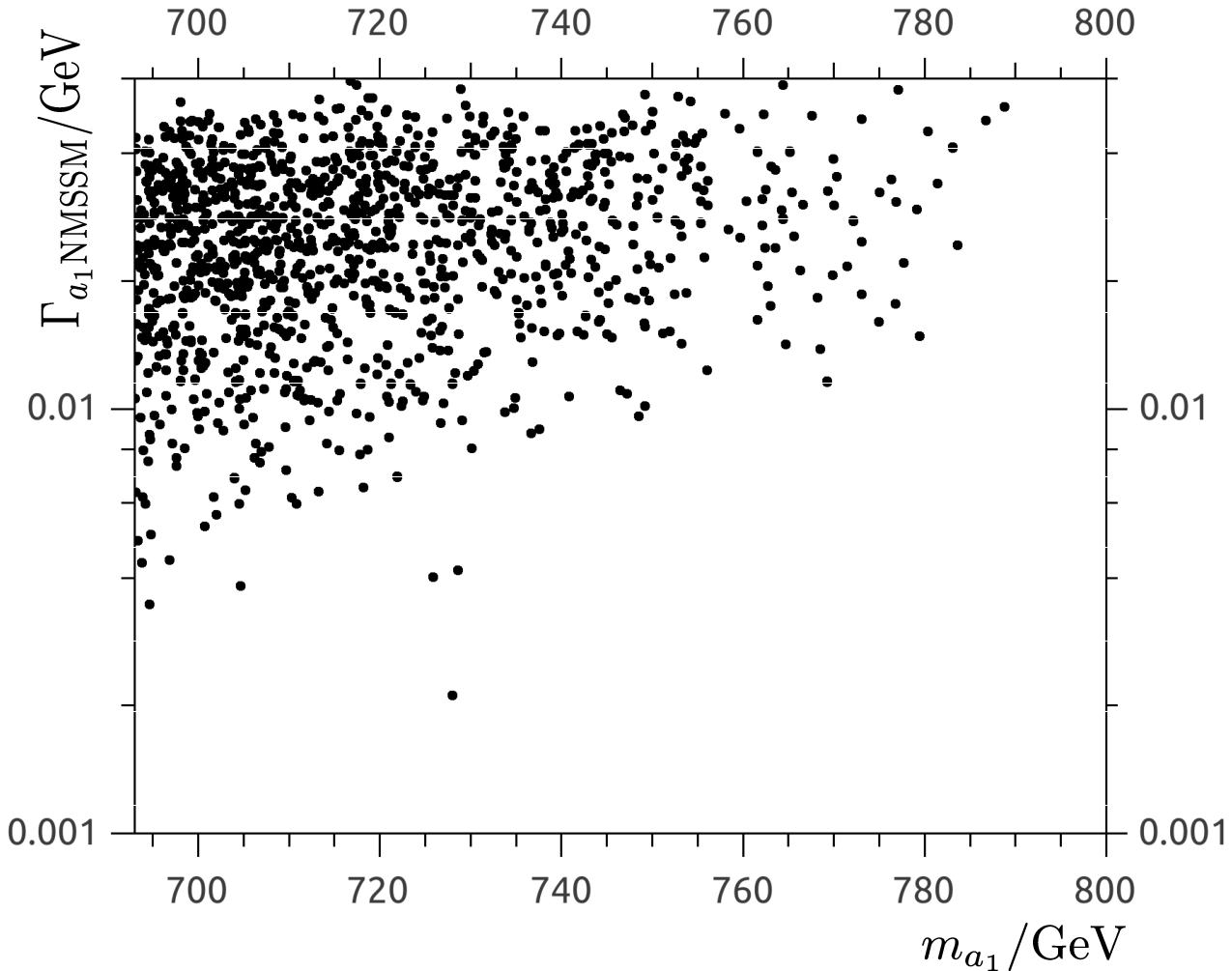}
\includegraphics[width=3in]{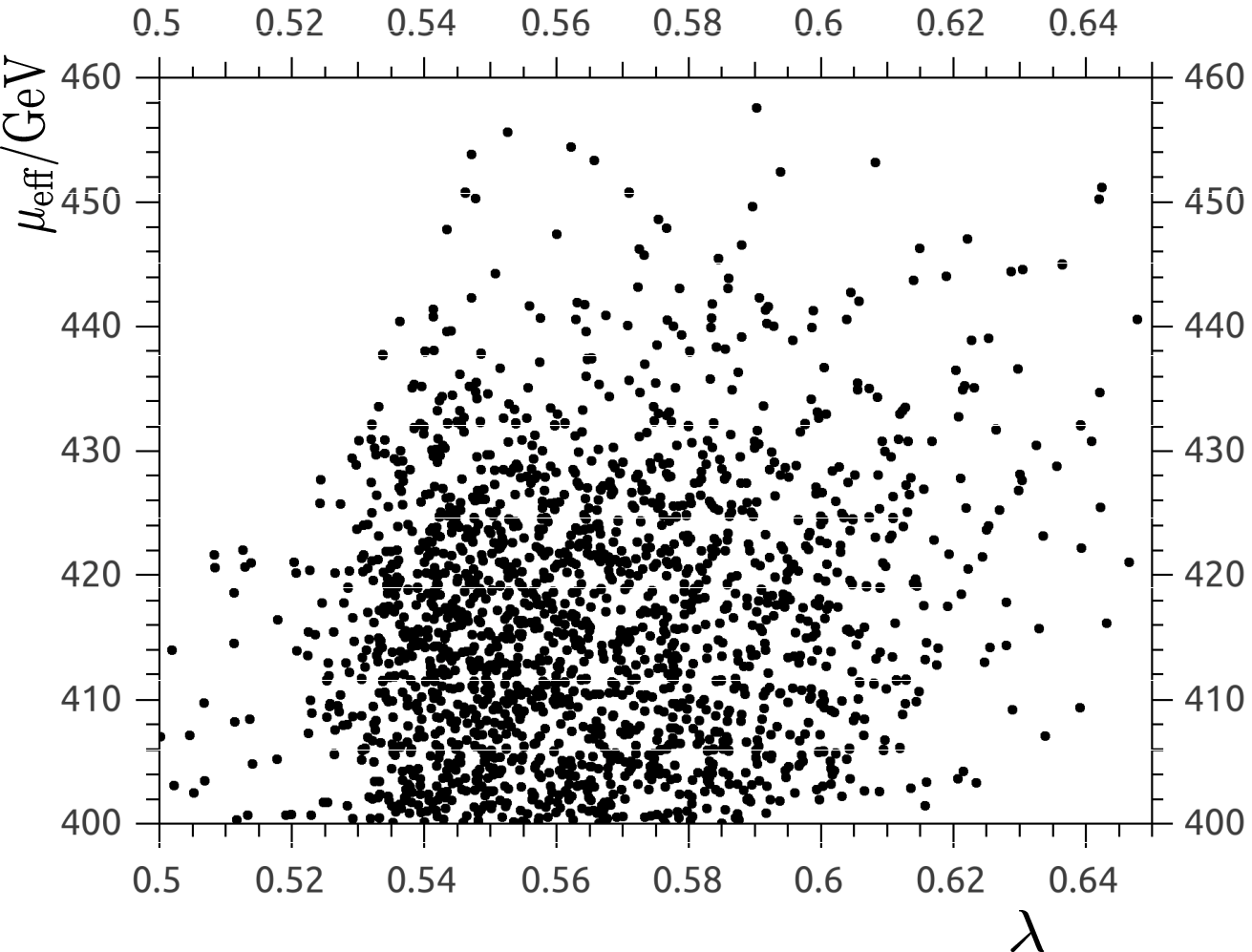}
\caption{The scanned points in the NMSSM parameter space, plotted on $m_{a_1}$-$\Gamma_{a_1 \text{NMSSM}}$ plane, and $\lambda$-$\mu_{\text{eff}}$ plane.} \label{NMSSMFigure}
\end{figure}

We plot the scanned points in the Fig.~\ref{NMSSMFigure}. We can see that there are at least some points that can reach $\Gamma_{a_1 \text{NMSSM}} \lesssim \Gamma_{10+\overline{10}}(a_1 \rightarrow g g), \Gamma_{3 \times (5+\overline{5})}(a_1 \rightarrow g g)$, lowering the signal strength a little, so the results in \ref{1010barSituation} are still not severely altered. Here we show a benchmark point,
\begin{eqnarray}
& & \lambda = 0.5471, ~~~\kappa = 0.3190, ~~~\mu_{\text{eff}} = 453.9 \text{ GeV}, ~~~\tan \beta = 2.060, \nonumber \\
& & A_{\lambda} = 589.6 \text{ GeV}, ~~~A_{\kappa}=-664.5 \text{ GeV}, \nonumber \\
& & m_{a_1} = 728.0 \text{ GeV}, ~~~\Gamma_{a_1, \text{NMSSM}} = 0.00211 \text{ GeV}.
\end{eqnarray}
In this point, the values of the $\lambda$ and the $\kappa$ approach the values in (\ref{Runned1010bar}), and the mass CP-odd singlet-like Higgs is near 750 GeV, while its pure NMSSM width is small compared with the vectorlike particle induced one.

\section{Conclusions and discussions}

In this paper, we find that the 
NMSSM with the additional $10+\overline{10}$ vectorlike particles can explain the possible diphoton excess recently revealed by the ATLAS and the CMS collaborations. From the aspect of the NMSSM, there is also enough parameter space to acquire a singlet-like CP-odd Higgs with the narrow-enough $\Gamma_{a_1\text{NMSSM}}$,which is the necessary condition to account for the di-photon excess.

In the Ref.~\cite{DiPhoton73}, the authors discussed the loop induced associated $W^+ W^-$, $ZZ$, $Z \gamma$ decay modes in the TABLE III, together with the relevant experimental bound on Page 2. In the NMSSM extended with the $10+\overline{10}$ model, $Q$, $\overline{Q}$, $U$, $\overline{U}$ mainly contribute to the loop, roughly succeeded in escaping the experimental bounds. This will be tested as more and more data are collected.

\begin{acknowledgements}

We would like to thank Ran Ding, Chen Zhang, Ying-Nan Mao, Yang Zhou, Mengchao Zhang for helpful discussions.  This work was supported in part by the Natural Science Foundation of China (Grants No.~11135003 and No.~11375014).

\end{acknowledgements}

\bibliography{DiPhotonNMSSMVL}

\begin{thebibliography}{138}
\expandafter\ifx\csname natexlab\endcsname\relax\def\natexlab#1{#1}\fi
\expandafter\ifx\csname bibnamefont\endcsname\relax
  \def\bibnamefont#1{#1}\fi
\expandafter\ifx\csname bibfnamefont\endcsname\relax
  \def\bibfnamefont#1{#1}\fi
\expandafter\ifx\csname citenamefont\endcsname\relax
  \def\citenamefont#1{#1}\fi
\expandafter\ifx\csname url\endcsname\relax
  \def\url#1{\texttt{#1}}\fi
\expandafter\ifx\csname urlprefix\endcsname\relax\def\urlprefix{URL }\fi
\providecommand{\bibinfo}[2]{#2}
\providecommand{\eprint}[2][]{\url{#2}}

\bibitem[{\citenamefont{Tang}(2014)}]{MyPaper}
\bibinfo{author}{\bibfnamefont{Y.-L.} \bibnamefont{Tang}},
  \bibinfo{journal}{Phys. Rev.} \textbf{\bibinfo{volume}{D90}},
  \bibinfo{pages}{075020} (\bibinfo{year}{2014}), \eprint{1409.5858}.

\bibitem[{\citenamefont{collaboration}(2015)}]{ATLASDiPhoton}
\bibinfo{author}{\bibfnamefont{T.~A.} \bibnamefont{collaboration}}
  (\bibinfo{year}{2015}), \eprint{ATLAS-CONF-2015-081}.

\bibitem[{\citenamefont{Collaboration}(2015)}]{CMSDiPhoton}
\bibinfo{author}{\bibfnamefont{C.}~\bibnamefont{Collaboration}}
  (\bibinfo{collaboration}{CMS}) (\bibinfo{year}{2015}),
  \eprint{CMS-PAS-EXO-15-004}.

\bibitem[{\citenamefont{Hall et~al.}(2015)\citenamefont{Hall, Harigaya, and
  Nomura}}]{DiPhoton1}
\bibinfo{author}{\bibfnamefont{L.~J.} \bibnamefont{Hall}},
  \bibinfo{author}{\bibfnamefont{K.}~\bibnamefont{Harigaya}}, \bibnamefont{and}
  \bibinfo{author}{\bibfnamefont{Y.}~\bibnamefont{Nomura}}
  (\bibinfo{year}{2015}), \eprint{1512.07904}.

\bibitem[{\citenamefont{Casas et~al.}(2015)\citenamefont{Casas, Espinosa, and
  Moreno}}]{DiPhoton2}
\bibinfo{author}{\bibfnamefont{J.~A.} \bibnamefont{Casas}},
  \bibinfo{author}{\bibfnamefont{J.~R.} \bibnamefont{Espinosa}},
  \bibnamefont{and} \bibinfo{author}{\bibfnamefont{J.~M.} \bibnamefont{Moreno}}
  (\bibinfo{year}{2015}), \eprint{1512.07895}.

\bibitem[{\citenamefont{Zhang and Zhou}(2015)}]{DiPhoton3}
\bibinfo{author}{\bibfnamefont{J.}~\bibnamefont{Zhang}} \bibnamefont{and}
  \bibinfo{author}{\bibfnamefont{S.}~\bibnamefont{Zhou}}
  (\bibinfo{year}{2015}), \eprint{1512.07889}.

\bibitem[{\citenamefont{Liu et~al.}(2015)\citenamefont{Liu, Wang, and
  Xue}}]{DiPhoton4}
\bibinfo{author}{\bibfnamefont{J.}~\bibnamefont{Liu}},
  \bibinfo{author}{\bibfnamefont{X.-P.} \bibnamefont{Wang}}, \bibnamefont{and}
  \bibinfo{author}{\bibfnamefont{W.}~\bibnamefont{Xue}} (\bibinfo{year}{2015}),
  \eprint{1512.07885}.

\bibitem[{\citenamefont{Cheung et~al.}(2015)\citenamefont{Cheung, Ko, Lee,
  Park, and Tseng}}]{DiPhoton5}
\bibinfo{author}{\bibfnamefont{K.}~\bibnamefont{Cheung}},
  \bibinfo{author}{\bibfnamefont{P.}~\bibnamefont{Ko}},
  \bibinfo{author}{\bibfnamefont{J.~S.} \bibnamefont{Lee}},
  \bibinfo{author}{\bibfnamefont{J.}~\bibnamefont{Park}}, \bibnamefont{and}
  \bibinfo{author}{\bibfnamefont{P.-Y.} \bibnamefont{Tseng}}
  (\bibinfo{year}{2015}), \eprint{1512.07853}.

\bibitem[{\citenamefont{Das and Rai}(2015)}]{DiPhoton6}
\bibinfo{author}{\bibfnamefont{K.}~\bibnamefont{Das}} \bibnamefont{and}
  \bibinfo{author}{\bibfnamefont{S.~K.} \bibnamefont{Rai}}
  (\bibinfo{year}{2015}), \eprint{1512.07789}.

\bibitem[{\citenamefont{Davoudiasl and Zhang}(2015)}]{DiPhoton7}
\bibinfo{author}{\bibfnamefont{H.}~\bibnamefont{Davoudiasl}} \bibnamefont{and}
  \bibinfo{author}{\bibfnamefont{C.}~\bibnamefont{Zhang}}
  (\bibinfo{year}{2015}), \eprint{1512.07672}.

\bibitem[{\citenamefont{Cvetič et~al.}(2015)\citenamefont{Cvetič, Halverson,
  and Langacker}}]{DiPhoton8}
\bibinfo{author}{\bibfnamefont{M.}~\bibnamefont{Cvetič}},
  \bibinfo{author}{\bibfnamefont{J.}~\bibnamefont{Halverson}},
  \bibnamefont{and} \bibinfo{author}{\bibfnamefont{P.}~\bibnamefont{Langacker}}
  (\bibinfo{year}{2015}), \eprint{1512.07622}.

\bibitem[{\citenamefont{Altmannshofer et~al.}(2015)\citenamefont{Altmannshofer,
  Galloway, Gori, Kagan, Martin, and Zupan}}]{DiPhoton9}
\bibinfo{author}{\bibfnamefont{W.}~\bibnamefont{Altmannshofer}},
  \bibinfo{author}{\bibfnamefont{J.}~\bibnamefont{Galloway}},
  \bibinfo{author}{\bibfnamefont{S.}~\bibnamefont{Gori}},
  \bibinfo{author}{\bibfnamefont{A.~L.} \bibnamefont{Kagan}},
  \bibinfo{author}{\bibfnamefont{A.}~\bibnamefont{Martin}}, \bibnamefont{and}
  \bibinfo{author}{\bibfnamefont{J.}~\bibnamefont{Zupan}}
  (\bibinfo{year}{2015}), \eprint{1512.07616}.

\bibitem[{\citenamefont{Chakraborty et~al.}(2015)\citenamefont{Chakraborty,
  Chakraborty, and Raychaudhuri}}]{DiPhoton10}
\bibinfo{author}{\bibfnamefont{S.}~\bibnamefont{Chakraborty}},
  \bibinfo{author}{\bibfnamefont{A.}~\bibnamefont{Chakraborty}},
  \bibnamefont{and}
  \bibinfo{author}{\bibfnamefont{S.}~\bibnamefont{Raychaudhuri}}
  (\bibinfo{year}{2015}), \eprint{1512.07527}.

\bibitem[{\citenamefont{Badziak}(2015)}]{DiPhoton11}
\bibinfo{author}{\bibfnamefont{M.}~\bibnamefont{Badziak}}
  (\bibinfo{year}{2015}), \eprint{1512.07497}.

\bibitem[{\citenamefont{Patel and Sharma}(2015)}]{DiPhoton12}
\bibinfo{author}{\bibfnamefont{K.~M.} \bibnamefont{Patel}} \bibnamefont{and}
  \bibinfo{author}{\bibfnamefont{P.}~\bibnamefont{Sharma}}
  (\bibinfo{year}{2015}), \eprint{1512.07468}.

\bibitem[{\citenamefont{Chabab et~al.}(2015)\citenamefont{Chabab,
  Capdequi-Peyranère, and Rahili}}]{DiPhoton13}
\bibinfo{author}{\bibfnamefont{M.}~\bibnamefont{Chabab}},
  \bibinfo{author}{\bibfnamefont{M.}~\bibnamefont{Capdequi-Peyranère}},
  \bibnamefont{and} \bibinfo{author}{\bibfnamefont{L.}~\bibnamefont{Rahili}}
  (\bibinfo{year}{2015}), \eprint{1512.07280}.

\bibitem[{\citenamefont{Huang et~al.}(2015{\natexlab{a}})\citenamefont{Huang,
  Tsai, and Yuan}}]{DiPhoton14}
\bibinfo{author}{\bibfnamefont{W.-C.} \bibnamefont{Huang}},
  \bibinfo{author}{\bibfnamefont{Y.-L.~S.} \bibnamefont{Tsai}},
  \bibnamefont{and} \bibinfo{author}{\bibfnamefont{T.-C.} \bibnamefont{Yuan}}
  (\bibinfo{year}{2015}{\natexlab{a}}), \eprint{1512.07268}.

\bibitem[{\citenamefont{Belyaev et~al.}(2015)\citenamefont{Belyaev,
  Cacciapaglia, Cai, Flacke, Parolini, and Serôdio}}]{DiPhoton15}
\bibinfo{author}{\bibfnamefont{A.}~\bibnamefont{Belyaev}},
  \bibinfo{author}{\bibfnamefont{G.}~\bibnamefont{Cacciapaglia}},
  \bibinfo{author}{\bibfnamefont{H.}~\bibnamefont{Cai}},
  \bibinfo{author}{\bibfnamefont{T.}~\bibnamefont{Flacke}},
  \bibinfo{author}{\bibfnamefont{A.}~\bibnamefont{Parolini}}, \bibnamefont{and}
  \bibinfo{author}{\bibfnamefont{H.}~\bibnamefont{Serôdio}}
  (\bibinfo{year}{2015}), \eprint{1512.07242}.

\bibitem[{\citenamefont{Dey et~al.}(2015)\citenamefont{Dey, Mohanty, and
  Tomar}}]{DiPhoton16}
\bibinfo{author}{\bibfnamefont{U.~K.} \bibnamefont{Dey}},
  \bibinfo{author}{\bibfnamefont{S.}~\bibnamefont{Mohanty}}, \bibnamefont{and}
  \bibinfo{author}{\bibfnamefont{G.}~\bibnamefont{Tomar}}
  (\bibinfo{year}{2015}), \eprint{1512.07212}.

\bibitem[{\citenamefont{Hernández and Nisandzic}(2015)}]{DiPhoton17}
\bibinfo{author}{\bibfnamefont{A.~E.~C.} \bibnamefont{Hernández}}
  \bibnamefont{and} \bibinfo{author}{\bibfnamefont{I.}~\bibnamefont{Nisandzic}}
  (\bibinfo{year}{2015}), \eprint{1512.07165}.

\bibitem[{\citenamefont{Murphy}(2015)}]{DiPhoton18}
\bibinfo{author}{\bibfnamefont{C.~W.} \bibnamefont{Murphy}}
  (\bibinfo{year}{2015}), \eprint{1512.06976}.

\bibitem[{\citenamefont{Dev and Teresi}(2015)}]{DiPhoton19}
\bibinfo{author}{\bibfnamefont{P.~S.~B.} \bibnamefont{Dev}} \bibnamefont{and}
  \bibinfo{author}{\bibfnamefont{D.}~\bibnamefont{Teresi}}
  (\bibinfo{year}{2015}), \eprint{1512.07243}.

\bibitem[{\citenamefont{Kulkarni}(2015)}]{DiPhoton20}
\bibinfo{author}{\bibfnamefont{K.}~\bibnamefont{Kulkarni}}
  (\bibinfo{year}{2015}), \eprint{1512.06836}.

\bibitem[{\citenamefont{Chala et~al.}(2015)\citenamefont{Chala, Duerr,
  Kahlhoefer, and Schmidt-Hoberg}}]{DiPhoton21}
\bibinfo{author}{\bibfnamefont{M.}~\bibnamefont{Chala}},
  \bibinfo{author}{\bibfnamefont{M.}~\bibnamefont{Duerr}},
  \bibinfo{author}{\bibfnamefont{F.}~\bibnamefont{Kahlhoefer}},
  \bibnamefont{and}
  \bibinfo{author}{\bibfnamefont{K.}~\bibnamefont{Schmidt-Hoberg}}
  (\bibinfo{year}{2015}), \eprint{1512.06833}.

\bibitem[{\citenamefont{Cline and Liu}(2015)}]{DiPhoton22}
\bibinfo{author}{\bibfnamefont{J.~M.} \bibnamefont{Cline}} \bibnamefont{and}
  \bibinfo{author}{\bibfnamefont{Z.}~\bibnamefont{Liu}} (\bibinfo{year}{2015}),
  \eprint{1512.06827}.

\bibitem[{\citenamefont{Berthier et~al.}(2015)\citenamefont{Berthier, Cline,
  Shepherd, and Trott}}]{DiPhoton23}
\bibinfo{author}{\bibfnamefont{L.}~\bibnamefont{Berthier}},
  \bibinfo{author}{\bibfnamefont{J.~M.} \bibnamefont{Cline}},
  \bibinfo{author}{\bibfnamefont{W.}~\bibnamefont{Shepherd}}, \bibnamefont{and}
  \bibinfo{author}{\bibfnamefont{M.}~\bibnamefont{Trott}}
  (\bibinfo{year}{2015}), \eprint{1512.06799}.

\bibitem[{\citenamefont{Kim et~al.}(2015{\natexlab{a}})\citenamefont{Kim,
  Rolbiecki, and de~Austri}}]{DiPhoton24}
\bibinfo{author}{\bibfnamefont{J.~S.} \bibnamefont{Kim}},
  \bibinfo{author}{\bibfnamefont{K.}~\bibnamefont{Rolbiecki}},
  \bibnamefont{and} \bibinfo{author}{\bibfnamefont{R.~R.}
  \bibnamefont{de~Austri}} (\bibinfo{year}{2015}{\natexlab{a}}),
  \eprint{1512.06797}.

\bibitem[{\citenamefont{Bi et~al.}(2015)\citenamefont{Bi, Xiang, Yin, and
  Yu}}]{DiPhoton25}
\bibinfo{author}{\bibfnamefont{X.-J.} \bibnamefont{Bi}},
  \bibinfo{author}{\bibfnamefont{Q.-F.} \bibnamefont{Xiang}},
  \bibinfo{author}{\bibfnamefont{P.-F.} \bibnamefont{Yin}}, \bibnamefont{and}
  \bibinfo{author}{\bibfnamefont{Z.-H.} \bibnamefont{Yu}}
  (\bibinfo{year}{2015}), \eprint{1512.06787}.

\bibitem[{\citenamefont{Heckman}(2015)}]{DiPhoton26}
\bibinfo{author}{\bibfnamefont{J.~J.} \bibnamefont{Heckman}}
  (\bibinfo{year}{2015}), \eprint{1512.06773}.

\bibitem[{\citenamefont{Huang et~al.}(2015{\natexlab{b}})\citenamefont{Huang,
  Li, Liu, and Wang}}]{DiPhoton27}
\bibinfo{author}{\bibfnamefont{F.~P.} \bibnamefont{Huang}},
  \bibinfo{author}{\bibfnamefont{C.~S.} \bibnamefont{Li}},
  \bibinfo{author}{\bibfnamefont{Z.~L.} \bibnamefont{Liu}}, \bibnamefont{and}
  \bibinfo{author}{\bibfnamefont{Y.}~\bibnamefont{Wang}}
  (\bibinfo{year}{2015}{\natexlab{b}}), \eprint{1512.06732}.

\bibitem[{\citenamefont{Cao et~al.}(2015{\natexlab{a}})\citenamefont{Cao, Han,
  Shang, Su, Yang, and Zhang}}]{DiPhoton28}
\bibinfo{author}{\bibfnamefont{J.}~\bibnamefont{Cao}},
  \bibinfo{author}{\bibfnamefont{C.}~\bibnamefont{Han}},
  \bibinfo{author}{\bibfnamefont{L.}~\bibnamefont{Shang}},
  \bibinfo{author}{\bibfnamefont{W.}~\bibnamefont{Su}},
  \bibinfo{author}{\bibfnamefont{J.~M.} \bibnamefont{Yang}}, \bibnamefont{and}
  \bibinfo{author}{\bibfnamefont{Y.}~\bibnamefont{Zhang}}
  (\bibinfo{year}{2015}{\natexlab{a}}), \eprint{1512.06728}.

\bibitem[{\citenamefont{Wang et~al.}(2015)\citenamefont{Wang, Wu, Yang, and
  Zhang}}]{DiPhoton29}
\bibinfo{author}{\bibfnamefont{F.}~\bibnamefont{Wang}},
  \bibinfo{author}{\bibfnamefont{L.}~\bibnamefont{Wu}},
  \bibinfo{author}{\bibfnamefont{J.~M.} \bibnamefont{Yang}}, \bibnamefont{and}
  \bibinfo{author}{\bibfnamefont{M.}~\bibnamefont{Zhang}}
  (\bibinfo{year}{2015}), \eprint{1512.06715}.

\bibitem[{\citenamefont{Antipin et~al.}(2015)\citenamefont{Antipin, Mojaza, and
  Sannino}}]{DiPhoton30}
\bibinfo{author}{\bibfnamefont{O.}~\bibnamefont{Antipin}},
  \bibinfo{author}{\bibfnamefont{M.}~\bibnamefont{Mojaza}}, \bibnamefont{and}
  \bibinfo{author}{\bibfnamefont{F.}~\bibnamefont{Sannino}}
  (\bibinfo{year}{2015}), \eprint{1512.06708}.

\bibitem[{\citenamefont{Hatanaka}(2015)}]{DiPhoton31}
\bibinfo{author}{\bibfnamefont{H.}~\bibnamefont{Hatanaka}}
  (\bibinfo{year}{2015}), \eprint{1512.06595}.

\bibitem[{\citenamefont{Ding et~al.}(2015)\citenamefont{Ding, Huang, Li, and
  Zhu}}]{DiPhoton32}
\bibinfo{author}{\bibfnamefont{R.}~\bibnamefont{Ding}},
  \bibinfo{author}{\bibfnamefont{L.}~\bibnamefont{Huang}},
  \bibinfo{author}{\bibfnamefont{T.}~\bibnamefont{Li}}, \bibnamefont{and}
  \bibinfo{author}{\bibfnamefont{B.}~\bibnamefont{Zhu}} (\bibinfo{year}{2015}),
  \eprint{1512.06560}.

\bibitem[{\citenamefont{Chakraborty and Kundu}(2015)}]{DiPhoton33}
\bibinfo{author}{\bibfnamefont{I.}~\bibnamefont{Chakraborty}} \bibnamefont{and}
  \bibinfo{author}{\bibfnamefont{A.}~\bibnamefont{Kundu}}
  (\bibinfo{year}{2015}), \eprint{1512.06508}.

\bibitem[{\citenamefont{Barducci et~al.}(2015)\citenamefont{Barducci, Goudelis,
  Kulkarni, and Sengupta}}]{DiPhoton34}
\bibinfo{author}{\bibfnamefont{D.}~\bibnamefont{Barducci}},
  \bibinfo{author}{\bibfnamefont{A.}~\bibnamefont{Goudelis}},
  \bibinfo{author}{\bibfnamefont{S.}~\bibnamefont{Kulkarni}}, \bibnamefont{and}
  \bibinfo{author}{\bibfnamefont{D.}~\bibnamefont{Sengupta}}
  (\bibinfo{year}{2015}), \eprint{1512.06842}.

\bibitem[{\citenamefont{Feng et~al.}(2015)\citenamefont{Feng, Li, Zhang, and
  Zhao}}]{DiPhoton35}
\bibinfo{author}{\bibfnamefont{T.-F.} \bibnamefont{Feng}},
  \bibinfo{author}{\bibfnamefont{X.-Q.} \bibnamefont{Li}},
  \bibinfo{author}{\bibfnamefont{H.-B.} \bibnamefont{Zhang}}, \bibnamefont{and}
  \bibinfo{author}{\bibfnamefont{S.-M.} \bibnamefont{Zhao}}
  (\bibinfo{year}{2015}), \eprint{1512.06696}.

\bibitem[{\citenamefont{Bardhan et~al.}(2015)\citenamefont{Bardhan, Bhatia,
  Chakraborty, Maitra, Raychaudhuri, and Samui}}]{DiPhoton36}
\bibinfo{author}{\bibfnamefont{D.}~\bibnamefont{Bardhan}},
  \bibinfo{author}{\bibfnamefont{D.}~\bibnamefont{Bhatia}},
  \bibinfo{author}{\bibfnamefont{A.}~\bibnamefont{Chakraborty}},
  \bibinfo{author}{\bibfnamefont{U.}~\bibnamefont{Maitra}},
  \bibinfo{author}{\bibfnamefont{S.}~\bibnamefont{Raychaudhuri}},
  \bibnamefont{and} \bibinfo{author}{\bibfnamefont{T.}~\bibnamefont{Samui}}
  (\bibinfo{year}{2015}), \eprint{1512.06674}.

\bibitem[{\citenamefont{Chang et~al.}(2015)\citenamefont{Chang, Cheung, and
  Lu}}]{DiPhoton37}
\bibinfo{author}{\bibfnamefont{J.}~\bibnamefont{Chang}},
  \bibinfo{author}{\bibfnamefont{K.}~\bibnamefont{Cheung}}, \bibnamefont{and}
  \bibinfo{author}{\bibfnamefont{C.-T.} \bibnamefont{Lu}}
  (\bibinfo{year}{2015}), \eprint{1512.06671}.

\bibitem[{\citenamefont{Luo et~al.}(2015)\citenamefont{Luo, Wang, Xu, Zhang,
  and Zhu}}]{DiPhoton38}
\bibinfo{author}{\bibfnamefont{M.-x.} \bibnamefont{Luo}},
  \bibinfo{author}{\bibfnamefont{K.}~\bibnamefont{Wang}},
  \bibinfo{author}{\bibfnamefont{T.}~\bibnamefont{Xu}},
  \bibinfo{author}{\bibfnamefont{L.}~\bibnamefont{Zhang}}, \bibnamefont{and}
  \bibinfo{author}{\bibfnamefont{G.}~\bibnamefont{Zhu}} (\bibinfo{year}{2015}),
  \eprint{1512.06670}.

\bibitem[{\citenamefont{Han et~al.}(2015{\natexlab{a}})\citenamefont{Han, Wang,
  and Zheng}}]{DiPhoton39}
\bibinfo{author}{\bibfnamefont{H.}~\bibnamefont{Han}},
  \bibinfo{author}{\bibfnamefont{S.}~\bibnamefont{Wang}}, \bibnamefont{and}
  \bibinfo{author}{\bibfnamefont{S.}~\bibnamefont{Zheng}}
  (\bibinfo{year}{2015}{\natexlab{a}}), \eprint{1512.06562}.

\bibitem[{\citenamefont{Dhuria and Goswami}(2015)}]{DiPhoton40}
\bibinfo{author}{\bibfnamefont{M.}~\bibnamefont{Dhuria}} \bibnamefont{and}
  \bibinfo{author}{\bibfnamefont{G.}~\bibnamefont{Goswami}}
  (\bibinfo{year}{2015}), \eprint{1512.06782}.

\bibitem[{\citenamefont{Chang}(2015)}]{DiPhoton41}
\bibinfo{author}{\bibfnamefont{S.}~\bibnamefont{Chang}} (\bibinfo{year}{2015}),
  \eprint{1512.06426}.

\bibitem[{\citenamefont{Han et~al.}(2015{\natexlab{b}})\citenamefont{Han, Lee,
  Park, and Sanz}}]{DiPhoton42}
\bibinfo{author}{\bibfnamefont{C.}~\bibnamefont{Han}},
  \bibinfo{author}{\bibfnamefont{H.~M.} \bibnamefont{Lee}},
  \bibinfo{author}{\bibfnamefont{M.}~\bibnamefont{Park}}, \bibnamefont{and}
  \bibinfo{author}{\bibfnamefont{V.}~\bibnamefont{Sanz}}
  (\bibinfo{year}{2015}{\natexlab{b}}), \eprint{1512.06376}.

\bibitem[{\citenamefont{Arun and Saha}(2015)}]{DiPhoton43}
\bibinfo{author}{\bibfnamefont{M.~T.} \bibnamefont{Arun}} \bibnamefont{and}
  \bibinfo{author}{\bibfnamefont{P.}~\bibnamefont{Saha}}
  (\bibinfo{year}{2015}), \eprint{1512.06335}.

\bibitem[{\citenamefont{Ringwald and Saikawa}(2015)}]{DiPhoton44}
\bibinfo{author}{\bibfnamefont{A.}~\bibnamefont{Ringwald}} \bibnamefont{and}
  \bibinfo{author}{\bibfnamefont{K.}~\bibnamefont{Saikawa}}
  (\bibinfo{year}{2015}), \eprint{1512.06436}.

\bibitem[{\citenamefont{Chao}(2015)}]{DiPhoton45}
\bibinfo{author}{\bibfnamefont{W.}~\bibnamefont{Chao}} (\bibinfo{year}{2015}),
  \eprint{1512.06297}.

\bibitem[{\citenamefont{Carpenter et~al.}(2015)\citenamefont{Carpenter,
  Colburn, and Goodman}}]{DiPhoton46}
\bibinfo{author}{\bibfnamefont{L.~M.} \bibnamefont{Carpenter}},
  \bibinfo{author}{\bibfnamefont{R.}~\bibnamefont{Colburn}}, \bibnamefont{and}
  \bibinfo{author}{\bibfnamefont{J.}~\bibnamefont{Goodman}}
  (\bibinfo{year}{2015}), \eprint{1512.06107}.

\bibitem[{\citenamefont{Megias et~al.}(2015)\citenamefont{Megias, Pujolas, and
  Quiros}}]{DiPhoton47}
\bibinfo{author}{\bibfnamefont{E.}~\bibnamefont{Megias}},
  \bibinfo{author}{\bibfnamefont{O.}~\bibnamefont{Pujolas}}, \bibnamefont{and}
  \bibinfo{author}{\bibfnamefont{M.}~\bibnamefont{Quiros}}
  (\bibinfo{year}{2015}), \eprint{1512.06106}.

\bibitem[{\citenamefont{Alves et~al.}(2015{\natexlab{a}})\citenamefont{Alves,
  Dias, and Sinha}}]{DiPhoton48}
\bibinfo{author}{\bibfnamefont{A.}~\bibnamefont{Alves}},
  \bibinfo{author}{\bibfnamefont{A.~G.} \bibnamefont{Dias}}, \bibnamefont{and}
  \bibinfo{author}{\bibfnamefont{K.}~\bibnamefont{Sinha}}
  (\bibinfo{year}{2015}{\natexlab{a}}), \eprint{1512.06091}.

\bibitem[{\citenamefont{Kim et~al.}(2015{\natexlab{b}})\citenamefont{Kim,
  Reuter, Rolbiecki, and de~Austri}}]{DiPhoton49}
\bibinfo{author}{\bibfnamefont{J.~S.} \bibnamefont{Kim}},
  \bibinfo{author}{\bibfnamefont{J.}~\bibnamefont{Reuter}},
  \bibinfo{author}{\bibfnamefont{K.}~\bibnamefont{Rolbiecki}},
  \bibnamefont{and} \bibinfo{author}{\bibfnamefont{R.~R.}
  \bibnamefont{de~Austri}} (\bibinfo{year}{2015}{\natexlab{b}}),
  \eprint{1512.06083}.

\bibitem[{\citenamefont{Ghosh et~al.}(2015)\citenamefont{Ghosh, Kundu, and
  Ray}}]{DiPhoton50}
\bibinfo{author}{\bibfnamefont{S.}~\bibnamefont{Ghosh}},
  \bibinfo{author}{\bibfnamefont{A.}~\bibnamefont{Kundu}}, \bibnamefont{and}
  \bibinfo{author}{\bibfnamefont{S.}~\bibnamefont{Ray}} (\bibinfo{year}{2015}),
  \eprint{1512.05786}.

\bibitem[{\citenamefont{Bai et~al.}(2015)\citenamefont{Bai, Berger, and
  Lu}}]{DiPhoton51}
\bibinfo{author}{\bibfnamefont{Y.}~\bibnamefont{Bai}},
  \bibinfo{author}{\bibfnamefont{J.}~\bibnamefont{Berger}}, \bibnamefont{and}
  \bibinfo{author}{\bibfnamefont{R.}~\bibnamefont{Lu}} (\bibinfo{year}{2015}),
  \eprint{1512.05779}.

\bibitem[{\citenamefont{Falkowski et~al.}(2015)\citenamefont{Falkowski, Slone,
  and Volansky}}]{DiPhoton52}
\bibinfo{author}{\bibfnamefont{A.}~\bibnamefont{Falkowski}},
  \bibinfo{author}{\bibfnamefont{O.}~\bibnamefont{Slone}}, \bibnamefont{and}
  \bibinfo{author}{\bibfnamefont{T.}~\bibnamefont{Volansky}}
  (\bibinfo{year}{2015}), \eprint{1512.05777}.

\bibitem[{\citenamefont{Csaki et~al.}(2015)\citenamefont{Csaki, Hubisz, and
  Terning}}]{DiPhoton53}
\bibinfo{author}{\bibfnamefont{C.}~\bibnamefont{Csaki}},
  \bibinfo{author}{\bibfnamefont{J.}~\bibnamefont{Hubisz}}, \bibnamefont{and}
  \bibinfo{author}{\bibfnamefont{J.}~\bibnamefont{Terning}}
  (\bibinfo{year}{2015}), \eprint{1512.05776}.

\bibitem[{\citenamefont{Chakrabortty et~al.}(2015)\citenamefont{Chakrabortty,
  Choudhury, Ghosh, Mondal, and Srivastava}}]{DiPhoton54}
\bibinfo{author}{\bibfnamefont{J.}~\bibnamefont{Chakrabortty}},
  \bibinfo{author}{\bibfnamefont{A.}~\bibnamefont{Choudhury}},
  \bibinfo{author}{\bibfnamefont{P.}~\bibnamefont{Ghosh}},
  \bibinfo{author}{\bibfnamefont{S.}~\bibnamefont{Mondal}}, \bibnamefont{and}
  \bibinfo{author}{\bibfnamefont{T.}~\bibnamefont{Srivastava}}
  (\bibinfo{year}{2015}), \eprint{1512.05767}.

\bibitem[{\citenamefont{Bian et~al.}(2015)\citenamefont{Bian, Chen, Liu, and
  Shu}}]{DiPhoton55}
\bibinfo{author}{\bibfnamefont{L.}~\bibnamefont{Bian}},
  \bibinfo{author}{\bibfnamefont{N.}~\bibnamefont{Chen}},
  \bibinfo{author}{\bibfnamefont{D.}~\bibnamefont{Liu}}, \bibnamefont{and}
  \bibinfo{author}{\bibfnamefont{J.}~\bibnamefont{Shu}} (\bibinfo{year}{2015}),
  \eprint{1512.05759}.

\bibitem[{\citenamefont{Curtin and Verhaaren}(2015)}]{DiPhoton56}
\bibinfo{author}{\bibfnamefont{D.}~\bibnamefont{Curtin}} \bibnamefont{and}
  \bibinfo{author}{\bibfnamefont{C.~B.} \bibnamefont{Verhaaren}}
  (\bibinfo{year}{2015}), \eprint{1512.05753}.

\bibitem[{\citenamefont{Chao et~al.}(2015)\citenamefont{Chao, Huo, and
  Yu}}]{DiPhoton57}
\bibinfo{author}{\bibfnamefont{W.}~\bibnamefont{Chao}},
  \bibinfo{author}{\bibfnamefont{R.}~\bibnamefont{Huo}}, \bibnamefont{and}
  \bibinfo{author}{\bibfnamefont{J.-H.} \bibnamefont{Yu}}
  (\bibinfo{year}{2015}), \eprint{1512.05738}.

\bibitem[{\citenamefont{Demidov and Gorbunov}(2015)}]{DiPhoton58}
\bibinfo{author}{\bibfnamefont{S.~V.} \bibnamefont{Demidov}} \bibnamefont{and}
  \bibinfo{author}{\bibfnamefont{D.~S.} \bibnamefont{Gorbunov}}
  (\bibinfo{year}{2015}), \eprint{1512.05723}.

\bibitem[{\citenamefont{No et~al.}(2015)\citenamefont{No, Sanz, and
  Setford}}]{DiPhoton59}
\bibinfo{author}{\bibfnamefont{J.~M.} \bibnamefont{No}},
  \bibinfo{author}{\bibfnamefont{V.}~\bibnamefont{Sanz}}, \bibnamefont{and}
  \bibinfo{author}{\bibfnamefont{J.}~\bibnamefont{Setford}}
  (\bibinfo{year}{2015}), \eprint{1512.05700}.

\bibitem[{\citenamefont{Becirevic et~al.}(2015)\citenamefont{Becirevic,
  Bertuzzo, Sumensari, and Funchal}}]{DiPhoton60}
\bibinfo{author}{\bibfnamefont{D.}~\bibnamefont{Becirevic}},
  \bibinfo{author}{\bibfnamefont{E.}~\bibnamefont{Bertuzzo}},
  \bibinfo{author}{\bibfnamefont{O.}~\bibnamefont{Sumensari}},
  \bibnamefont{and} \bibinfo{author}{\bibfnamefont{R.~Z.}
  \bibnamefont{Funchal}} (\bibinfo{year}{2015}), \eprint{1512.05623}.

\bibitem[{\citenamefont{Agrawal et~al.}(2015)\citenamefont{Agrawal, Fan,
  Heidenreich, Reece, and Strassler}}]{DiPhoton61}
\bibinfo{author}{\bibfnamefont{P.}~\bibnamefont{Agrawal}},
  \bibinfo{author}{\bibfnamefont{J.}~\bibnamefont{Fan}},
  \bibinfo{author}{\bibfnamefont{B.}~\bibnamefont{Heidenreich}},
  \bibinfo{author}{\bibfnamefont{M.}~\bibnamefont{Reece}}, \bibnamefont{and}
  \bibinfo{author}{\bibfnamefont{M.}~\bibnamefont{Strassler}}
  (\bibinfo{year}{2015}), \eprint{1512.05775}.

\bibitem[{\citenamefont{Ahmed et~al.}(2015)\citenamefont{Ahmed, Dillon,
  Grzadkowski, Gunion, and Jiang}}]{DiPhoton62}
\bibinfo{author}{\bibfnamefont{A.}~\bibnamefont{Ahmed}},
  \bibinfo{author}{\bibfnamefont{B.~M.} \bibnamefont{Dillon}},
  \bibinfo{author}{\bibfnamefont{B.}~\bibnamefont{Grzadkowski}},
  \bibinfo{author}{\bibfnamefont{J.~F.} \bibnamefont{Gunion}},
  \bibnamefont{and} \bibinfo{author}{\bibfnamefont{Y.}~\bibnamefont{Jiang}}
  (\bibinfo{year}{2015}), \eprint{1512.05771}.

\bibitem[{\citenamefont{Cox et~al.}(2015)\citenamefont{Cox, Medina, Ray, and
  Spray}}]{DiPhoton63}
\bibinfo{author}{\bibfnamefont{P.}~\bibnamefont{Cox}},
  \bibinfo{author}{\bibfnamefont{A.~D.} \bibnamefont{Medina}},
  \bibinfo{author}{\bibfnamefont{T.~S.} \bibnamefont{Ray}}, \bibnamefont{and}
  \bibinfo{author}{\bibfnamefont{A.}~\bibnamefont{Spray}}
  (\bibinfo{year}{2015}), \eprint{1512.05618}.

\bibitem[{\citenamefont{Kobakhidze et~al.}(2015)\citenamefont{Kobakhidze, Wang,
  Wu, Yang, and Zhang}}]{DiPhoton64}
\bibinfo{author}{\bibfnamefont{A.}~\bibnamefont{Kobakhidze}},
  \bibinfo{author}{\bibfnamefont{F.}~\bibnamefont{Wang}},
  \bibinfo{author}{\bibfnamefont{L.}~\bibnamefont{Wu}},
  \bibinfo{author}{\bibfnamefont{J.~M.} \bibnamefont{Yang}}, \bibnamefont{and}
  \bibinfo{author}{\bibfnamefont{M.}~\bibnamefont{Zhang}}
  (\bibinfo{year}{2015}), \eprint{1512.05585}.

\bibitem[{\citenamefont{Matsuzaki and Yamawaki}(2015)}]{DiPhoton65}
\bibinfo{author}{\bibfnamefont{S.}~\bibnamefont{Matsuzaki}} \bibnamefont{and}
  \bibinfo{author}{\bibfnamefont{K.}~\bibnamefont{Yamawaki}}
  (\bibinfo{year}{2015}), \eprint{1512.05564}.

\bibitem[{\citenamefont{Cao et~al.}(2015{\natexlab{b}})\citenamefont{Cao, Liu,
  Xie, Yan, and Zhang}}]{DiPhoton66}
\bibinfo{author}{\bibfnamefont{Q.-H.} \bibnamefont{Cao}},
  \bibinfo{author}{\bibfnamefont{Y.}~\bibnamefont{Liu}},
  \bibinfo{author}{\bibfnamefont{K.-P.} \bibnamefont{Xie}},
  \bibinfo{author}{\bibfnamefont{B.}~\bibnamefont{Yan}}, \bibnamefont{and}
  \bibinfo{author}{\bibfnamefont{D.-M.} \bibnamefont{Zhang}}
  (\bibinfo{year}{2015}{\natexlab{b}}), \eprint{1512.05542}.

\bibitem[{\citenamefont{Dutta et~al.}(2015)\citenamefont{Dutta, Gao, Ghosh,
  Gogoladze, and Li}}]{DiPhoton67}
\bibinfo{author}{\bibfnamefont{B.}~\bibnamefont{Dutta}},
  \bibinfo{author}{\bibfnamefont{Y.}~\bibnamefont{Gao}},
  \bibinfo{author}{\bibfnamefont{T.}~\bibnamefont{Ghosh}},
  \bibinfo{author}{\bibfnamefont{I.}~\bibnamefont{Gogoladze}},
  \bibnamefont{and} \bibinfo{author}{\bibfnamefont{T.}~\bibnamefont{Li}}
  (\bibinfo{year}{2015}), \eprint{1512.05439}.

\bibitem[{\citenamefont{Petersson and Torre}(2015)}]{DiPhoton68}
\bibinfo{author}{\bibfnamefont{C.}~\bibnamefont{Petersson}} \bibnamefont{and}
  \bibinfo{author}{\bibfnamefont{R.}~\bibnamefont{Torre}}
  (\bibinfo{year}{2015}), \eprint{1512.05333}.

\bibitem[{\citenamefont{Low et~al.}(2015)\citenamefont{Low, Tesi, and
  Wang}}]{DiPhoton69}
\bibinfo{author}{\bibfnamefont{M.}~\bibnamefont{Low}},
  \bibinfo{author}{\bibfnamefont{A.}~\bibnamefont{Tesi}}, \bibnamefont{and}
  \bibinfo{author}{\bibfnamefont{L.-T.} \bibnamefont{Wang}}
  (\bibinfo{year}{2015}), \eprint{1512.05328}.

\bibitem[{\citenamefont{Molinaro et~al.}(2015)\citenamefont{Molinaro, Sannino,
  and Vignaroli}}]{DiPhoton70}
\bibinfo{author}{\bibfnamefont{E.}~\bibnamefont{Molinaro}},
  \bibinfo{author}{\bibfnamefont{F.}~\bibnamefont{Sannino}}, \bibnamefont{and}
  \bibinfo{author}{\bibfnamefont{N.}~\bibnamefont{Vignaroli}}
  (\bibinfo{year}{2015}), \eprint{1512.05334}.

\bibitem[{\citenamefont{Gupta et~al.}(2015)\citenamefont{Gupta, Jäger, Kats,
  Perez, and Stamou}}]{DiPhoton71}
\bibinfo{author}{\bibfnamefont{R.~S.} \bibnamefont{Gupta}},
  \bibinfo{author}{\bibfnamefont{S.}~\bibnamefont{Jäger}},
  \bibinfo{author}{\bibfnamefont{Y.}~\bibnamefont{Kats}},
  \bibinfo{author}{\bibfnamefont{G.}~\bibnamefont{Perez}}, \bibnamefont{and}
  \bibinfo{author}{\bibfnamefont{E.}~\bibnamefont{Stamou}}
  (\bibinfo{year}{2015}), \eprint{1512.05332}.

\bibitem[{\citenamefont{Ellis et~al.}(2015)\citenamefont{Ellis, Ellis,
  Quevillon, Sanz, and You}}]{DiPhoton72}
\bibinfo{author}{\bibfnamefont{J.}~\bibnamefont{Ellis}},
  \bibinfo{author}{\bibfnamefont{S.~A.~R.} \bibnamefont{Ellis}},
  \bibinfo{author}{\bibfnamefont{J.}~\bibnamefont{Quevillon}},
  \bibinfo{author}{\bibfnamefont{V.}~\bibnamefont{Sanz}}, \bibnamefont{and}
  \bibinfo{author}{\bibfnamefont{T.}~\bibnamefont{You}} (\bibinfo{year}{2015}),
  \eprint{1512.05327}.

\bibitem[{\citenamefont{Knapen et~al.}(2015)\citenamefont{Knapen, Melia,
  Papucci, and Zurek}}]{DiPhoton73}
\bibinfo{author}{\bibfnamefont{S.}~\bibnamefont{Knapen}},
  \bibinfo{author}{\bibfnamefont{T.}~\bibnamefont{Melia}},
  \bibinfo{author}{\bibfnamefont{M.}~\bibnamefont{Papucci}}, \bibnamefont{and}
  \bibinfo{author}{\bibfnamefont{K.}~\bibnamefont{Zurek}}
  (\bibinfo{year}{2015}), \eprint{1512.04928}.

\bibitem[{\citenamefont{Buttazzo et~al.}(2015)\citenamefont{Buttazzo, Greljo,
  and Marzocca}}]{DiPhoton74}
\bibinfo{author}{\bibfnamefont{D.}~\bibnamefont{Buttazzo}},
  \bibinfo{author}{\bibfnamefont{A.}~\bibnamefont{Greljo}}, \bibnamefont{and}
  \bibinfo{author}{\bibfnamefont{D.}~\bibnamefont{Marzocca}}
  (\bibinfo{year}{2015}), \eprint{1512.04929}.

\bibitem[{\citenamefont{Nakai et~al.}(2015)\citenamefont{Nakai, Sato, and
  Tobioka}}]{DiPhoton75}
\bibinfo{author}{\bibfnamefont{Y.}~\bibnamefont{Nakai}},
  \bibinfo{author}{\bibfnamefont{R.}~\bibnamefont{Sato}}, \bibnamefont{and}
  \bibinfo{author}{\bibfnamefont{K.}~\bibnamefont{Tobioka}}
  (\bibinfo{year}{2015}), \eprint{1512.04924}.

\bibitem[{\citenamefont{Harigaya and Nomura}(2015)}]{DiPhoton76}
\bibinfo{author}{\bibfnamefont{K.}~\bibnamefont{Harigaya}} \bibnamefont{and}
  \bibinfo{author}{\bibfnamefont{Y.}~\bibnamefont{Nomura}}
  (\bibinfo{year}{2015}), \eprint{1512.04850}.

\bibitem[{\citenamefont{Mambrini et~al.}(2015)\citenamefont{Mambrini, Arcadi,
  and Djouadi}}]{DiPhotonAdded1}
\bibinfo{author}{\bibfnamefont{Y.}~\bibnamefont{Mambrini}},
  \bibinfo{author}{\bibfnamefont{G.}~\bibnamefont{Arcadi}}, \bibnamefont{and}
  \bibinfo{author}{\bibfnamefont{A.}~\bibnamefont{Djouadi}}
  (\bibinfo{year}{2015}), \eprint{1512.04913}.

\bibitem[{\citenamefont{Backovic et~al.}(2015)\citenamefont{Backovic, Mariotti,
  and Redigolo}}]{DiPhotonAdded2}
\bibinfo{author}{\bibfnamefont{M.}~\bibnamefont{Backovic}},
  \bibinfo{author}{\bibfnamefont{A.}~\bibnamefont{Mariotti}}, \bibnamefont{and}
  \bibinfo{author}{\bibfnamefont{D.}~\bibnamefont{Redigolo}}
  (\bibinfo{year}{2015}), \eprint{1512.04917}.

\bibitem[{\citenamefont{Angelescu
  et~al.}(2015{\natexlab{a}})\citenamefont{Angelescu, Djouadi, and
  Moreau}}]{DiPhotonAdded3}
\bibinfo{author}{\bibfnamefont{A.}~\bibnamefont{Angelescu}},
  \bibinfo{author}{\bibfnamefont{A.}~\bibnamefont{Djouadi}}, \bibnamefont{and}
  \bibinfo{author}{\bibfnamefont{G.}~\bibnamefont{Moreau}}
  (\bibinfo{year}{2015}{\natexlab{a}}), \eprint{1512.04921}.

\bibitem[{\citenamefont{Pilaftsis}(2015)}]{DiPhotonAdded4}
\bibinfo{author}{\bibfnamefont{A.}~\bibnamefont{Pilaftsis}}
  (\bibinfo{year}{2015}), \eprint{1512.04931}.

\bibitem[{\citenamefont{Franceschini et~al.}(2015)\citenamefont{Franceschini,
  Giudice, Kamenik, McCullough, Pomarol, Rattazzi, Redi, Riva, Strumia, and
  Torre}}]{DiPhotonAdded5}
\bibinfo{author}{\bibfnamefont{R.}~\bibnamefont{Franceschini}},
  \bibinfo{author}{\bibfnamefont{G.~F.} \bibnamefont{Giudice}},
  \bibinfo{author}{\bibfnamefont{J.~F.} \bibnamefont{Kamenik}},
  \bibinfo{author}{\bibfnamefont{M.}~\bibnamefont{McCullough}},
  \bibinfo{author}{\bibfnamefont{A.}~\bibnamefont{Pomarol}},
  \bibinfo{author}{\bibfnamefont{R.}~\bibnamefont{Rattazzi}},
  \bibinfo{author}{\bibfnamefont{M.}~\bibnamefont{Redi}},
  \bibinfo{author}{\bibfnamefont{F.}~\bibnamefont{Riva}},
  \bibinfo{author}{\bibfnamefont{A.}~\bibnamefont{Strumia}}, \bibnamefont{and}
  \bibinfo{author}{\bibfnamefont{R.}~\bibnamefont{Torre}}
  (\bibinfo{year}{2015}), \eprint{1512.04933}.

\bibitem[{\citenamefont{Di~Chiara et~al.}(2015)\citenamefont{Di~Chiara,
  Marzola, and Raidal}}]{DiPhotonAdded6}
\bibinfo{author}{\bibfnamefont{S.}~\bibnamefont{Di~Chiara}},
  \bibinfo{author}{\bibfnamefont{L.}~\bibnamefont{Marzola}}, \bibnamefont{and}
  \bibinfo{author}{\bibfnamefont{M.}~\bibnamefont{Raidal}}
  (\bibinfo{year}{2015}), \eprint{1512.04939}.

\bibitem[{\citenamefont{Higaki et~al.}(2015)\citenamefont{Higaki, Jeong,
  Kitajima, and Takahashi}}]{DiPhotonAdded7}
\bibinfo{author}{\bibfnamefont{T.}~\bibnamefont{Higaki}},
  \bibinfo{author}{\bibfnamefont{K.~S.} \bibnamefont{Jeong}},
  \bibinfo{author}{\bibfnamefont{N.}~\bibnamefont{Kitajima}}, \bibnamefont{and}
  \bibinfo{author}{\bibfnamefont{F.}~\bibnamefont{Takahashi}}
  (\bibinfo{year}{2015}), \eprint{1512.05295}.

\bibitem[{\citenamefont{McDermott et~al.}(2015)\citenamefont{McDermott, Meade,
  and Ramani}}]{DiPhotonAdded8}
\bibinfo{author}{\bibfnamefont{S.~D.} \bibnamefont{McDermott}},
  \bibinfo{author}{\bibfnamefont{P.}~\bibnamefont{Meade}}, \bibnamefont{and}
  \bibinfo{author}{\bibfnamefont{H.}~\bibnamefont{Ramani}}
  (\bibinfo{year}{2015}), \eprint{1512.05326}.

\bibitem[{\citenamefont{Bellazzini et~al.}(2015)\citenamefont{Bellazzini,
  Franceschini, Sala, and Serra}}]{DiPhotonAdded9}
\bibinfo{author}{\bibfnamefont{B.}~\bibnamefont{Bellazzini}},
  \bibinfo{author}{\bibfnamefont{R.}~\bibnamefont{Franceschini}},
  \bibinfo{author}{\bibfnamefont{F.}~\bibnamefont{Sala}}, \bibnamefont{and}
  \bibinfo{author}{\bibfnamefont{J.}~\bibnamefont{Serra}}
  (\bibinfo{year}{2015}), \eprint{1512.05330}.

\bibitem[{\citenamefont{Martinez et~al.}(2015)\citenamefont{Martinez, Ochoa,
  and Sierra}}]{DiPhotonAdded10}
\bibinfo{author}{\bibfnamefont{R.}~\bibnamefont{Martinez}},
  \bibinfo{author}{\bibfnamefont{F.}~\bibnamefont{Ochoa}}, \bibnamefont{and}
  \bibinfo{author}{\bibfnamefont{C.~F.} \bibnamefont{Sierra}}
  (\bibinfo{year}{2015}), \eprint{1512.05617}.

\bibitem[{\citenamefont{Fichet et~al.}(2015)\citenamefont{Fichet, von
  Gersdorff, and Royon}}]{DiPhotonAdded11}
\bibinfo{author}{\bibfnamefont{S.}~\bibnamefont{Fichet}},
  \bibinfo{author}{\bibfnamefont{G.}~\bibnamefont{von Gersdorff}},
  \bibnamefont{and} \bibinfo{author}{\bibfnamefont{C.}~\bibnamefont{Royon}}
  (\bibinfo{year}{2015}), \eprint{1512.05751}.

\bibitem[{\citenamefont{Aloni et~al.}(2015)\citenamefont{Aloni, Blum, Dery,
  Efrati, and Nir}}]{DiPhotonAdded12}
\bibinfo{author}{\bibfnamefont{D.}~\bibnamefont{Aloni}},
  \bibinfo{author}{\bibfnamefont{K.}~\bibnamefont{Blum}},
  \bibinfo{author}{\bibfnamefont{A.}~\bibnamefont{Dery}},
  \bibinfo{author}{\bibfnamefont{A.}~\bibnamefont{Efrati}}, \bibnamefont{and}
  \bibinfo{author}{\bibfnamefont{Y.}~\bibnamefont{Nir}} (\bibinfo{year}{2015}),
  \eprint{1512.05778}.

\bibitem[{\citenamefont{Gabrielli et~al.}(2015)\citenamefont{Gabrielli,
  Kannike, Mele, Raidal, Spethmann, and Veermäe}}]{DiPhotonAdded13}
\bibinfo{author}{\bibfnamefont{E.}~\bibnamefont{Gabrielli}},
  \bibinfo{author}{\bibfnamefont{K.}~\bibnamefont{Kannike}},
  \bibinfo{author}{\bibfnamefont{B.}~\bibnamefont{Mele}},
  \bibinfo{author}{\bibfnamefont{M.}~\bibnamefont{Raidal}},
  \bibinfo{author}{\bibfnamefont{C.}~\bibnamefont{Spethmann}},
  \bibnamefont{and} \bibinfo{author}{\bibfnamefont{H.}~\bibnamefont{Veermäe}}
  (\bibinfo{year}{2015}), \eprint{1512.05961}.

\bibitem[{\citenamefont{Benbrik et~al.}(2015)\citenamefont{Benbrik, Chen, and
  Nomura}}]{DiPhotonAdded14}
\bibinfo{author}{\bibfnamefont{R.}~\bibnamefont{Benbrik}},
  \bibinfo{author}{\bibfnamefont{C.-H.} \bibnamefont{Chen}}, \bibnamefont{and}
  \bibinfo{author}{\bibfnamefont{T.}~\bibnamefont{Nomura}}
  (\bibinfo{year}{2015}), \eprint{1512.06028}.

\bibitem[{\citenamefont{Bernon and Smith}(2015)}]{DiPhotonAdded15}
\bibinfo{author}{\bibfnamefont{J.}~\bibnamefont{Bernon}} \bibnamefont{and}
  \bibinfo{author}{\bibfnamefont{C.}~\bibnamefont{Smith}}
  (\bibinfo{year}{2015}), \eprint{1512.06113}.

\bibitem[{\citenamefont{Han and Wang}(2015)}]{DiPhotonAdded16}
\bibinfo{author}{\bibfnamefont{X.-F.} \bibnamefont{Han}} \bibnamefont{and}
  \bibinfo{author}{\bibfnamefont{L.}~\bibnamefont{Wang}}
  (\bibinfo{year}{2015}), \eprint{1512.06587}.

\bibitem[{\citenamefont{Liao and Zheng}(2015)}]{DiPhotonAdded17}
\bibinfo{author}{\bibfnamefont{W.}~\bibnamefont{Liao}} \bibnamefont{and}
  \bibinfo{author}{\bibfnamefont{H.-q.} \bibnamefont{Zheng}}
  (\bibinfo{year}{2015}), \eprint{1512.06741}.

\bibitem[{\citenamefont{Cho et~al.}(2015)\citenamefont{Cho, Kim, Kong, Lim,
  Matchev, Park, and Park}}]{DiPhotonAdded18}
\bibinfo{author}{\bibfnamefont{W.~S.} \bibnamefont{Cho}},
  \bibinfo{author}{\bibfnamefont{D.}~\bibnamefont{Kim}},
  \bibinfo{author}{\bibfnamefont{K.}~\bibnamefont{Kong}},
  \bibinfo{author}{\bibfnamefont{S.~H.} \bibnamefont{Lim}},
  \bibinfo{author}{\bibfnamefont{K.~T.} \bibnamefont{Matchev}},
  \bibinfo{author}{\bibfnamefont{J.-C.} \bibnamefont{Park}}, \bibnamefont{and}
  \bibinfo{author}{\bibfnamefont{M.}~\bibnamefont{Park}}
  (\bibinfo{year}{2015}), \eprint{1512.06824}.

\bibitem[{\citenamefont{Bauer and Neubert}(2015)}]{DiPhotonAdded19}
\bibinfo{author}{\bibfnamefont{M.}~\bibnamefont{Bauer}} \bibnamefont{and}
  \bibinfo{author}{\bibfnamefont{M.}~\bibnamefont{Neubert}}
  (\bibinfo{year}{2015}), \eprint{1512.06828}.

\bibitem[{\citenamefont{Boucenna et~al.}(2015)\citenamefont{Boucenna, Morisi,
  and Vicente}}]{DiPhotonAdded20}
\bibinfo{author}{\bibfnamefont{S.~M.} \bibnamefont{Boucenna}},
  \bibinfo{author}{\bibfnamefont{S.}~\bibnamefont{Morisi}}, \bibnamefont{and}
  \bibinfo{author}{\bibfnamefont{A.}~\bibnamefont{Vicente}}
  (\bibinfo{year}{2015}), \eprint{1512.06878}.

\bibitem[{\citenamefont{Pelaggi et~al.}(2015)\citenamefont{Pelaggi, Strumia,
  and Vigiani}}]{DiPhotonAdded21}
\bibinfo{author}{\bibfnamefont{G.~M.} \bibnamefont{Pelaggi}},
  \bibinfo{author}{\bibfnamefont{A.}~\bibnamefont{Strumia}}, \bibnamefont{and}
  \bibinfo{author}{\bibfnamefont{E.}~\bibnamefont{Vigiani}}
  (\bibinfo{year}{2015}), \eprint{1512.07225}.

\bibitem[{\citenamefont{de~Blas et~al.}(2015)\citenamefont{de~Blas, Santiago,
  and Vega-Morales}}]{DiPhotonAdded22}
\bibinfo{author}{\bibfnamefont{J.}~\bibnamefont{de~Blas}},
  \bibinfo{author}{\bibfnamefont{J.}~\bibnamefont{Santiago}}, \bibnamefont{and}
  \bibinfo{author}{\bibfnamefont{R.}~\bibnamefont{Vega-Morales}}
  (\bibinfo{year}{2015}), \eprint{1512.07229}.

\bibitem[{\citenamefont{Moretti and Yagyu}(2015)}]{DiPhotonAdded23}
\bibinfo{author}{\bibfnamefont{S.}~\bibnamefont{Moretti}} \bibnamefont{and}
  \bibinfo{author}{\bibfnamefont{K.}~\bibnamefont{Yagyu}}
  (\bibinfo{year}{2015}), \eprint{1512.07462}.

\bibitem[{\citenamefont{Cao et~al.}(2015{\natexlab{c}})\citenamefont{Cao, Chen,
  and Gu}}]{DiPhotonAdded24}
\bibinfo{author}{\bibfnamefont{Q.-H.} \bibnamefont{Cao}},
  \bibinfo{author}{\bibfnamefont{S.-L.} \bibnamefont{Chen}}, \bibnamefont{and}
  \bibinfo{author}{\bibfnamefont{P.-H.} \bibnamefont{Gu}}
  (\bibinfo{year}{2015}{\natexlab{c}}), \eprint{1512.07541}.

\bibitem[{\citenamefont{Gu and Liu}(2015)}]{DiPhotonAdded25}
\bibinfo{author}{\bibfnamefont{J.}~\bibnamefont{Gu}} \bibnamefont{and}
  \bibinfo{author}{\bibfnamefont{Z.}~\bibnamefont{Liu}} (\bibinfo{year}{2015}),
  \eprint{1512.07624}.

\bibitem[{\citenamefont{Allanach et~al.}(2015)\citenamefont{Allanach, Dev,
  Renner, and Sakurai}}]{DiPhotonAdded26}
\bibinfo{author}{\bibfnamefont{B.~C.} \bibnamefont{Allanach}},
  \bibinfo{author}{\bibfnamefont{P.~S.~B.} \bibnamefont{Dev}},
  \bibinfo{author}{\bibfnamefont{S.~A.} \bibnamefont{Renner}},
  \bibnamefont{and} \bibinfo{author}{\bibfnamefont{K.}~\bibnamefont{Sakurai}}
  (\bibinfo{year}{2015}), \eprint{1512.07645}.

\bibitem[{\citenamefont{Craig et~al.}(2015)\citenamefont{Craig, Draper, Kilic,
  and Thomas}}]{DiPhotonAdded27}
\bibinfo{author}{\bibfnamefont{N.}~\bibnamefont{Craig}},
  \bibinfo{author}{\bibfnamefont{P.}~\bibnamefont{Draper}},
  \bibinfo{author}{\bibfnamefont{C.}~\bibnamefont{Kilic}}, \bibnamefont{and}
  \bibinfo{author}{\bibfnamefont{S.}~\bibnamefont{Thomas}}
  (\bibinfo{year}{2015}), \eprint{1512.07733}.

\bibitem[{\citenamefont{Han et~al.}(2015{\natexlab{c}})\citenamefont{Han, Wang,
  and Zheng}}]{DiPhotonAdded28}
\bibinfo{author}{\bibfnamefont{H.}~\bibnamefont{Han}},
  \bibinfo{author}{\bibfnamefont{S.}~\bibnamefont{Wang}}, \bibnamefont{and}
  \bibinfo{author}{\bibfnamefont{S.}~\bibnamefont{Zheng}}
  (\bibinfo{year}{2015}{\natexlab{c}}), \eprint{1512.07992}.

\bibitem[{\citenamefont{Hamada et~al.}(2015)\citenamefont{Hamada, Noumi, Sun,
  and Shiu}}]{ForceToAdd}
\bibinfo{author}{\bibfnamefont{Y.}~\bibnamefont{Hamada}},
  \bibinfo{author}{\bibfnamefont{T.}~\bibnamefont{Noumi}},
  \bibinfo{author}{\bibfnamefont{S.}~\bibnamefont{Sun}}, \bibnamefont{and}
  \bibinfo{author}{\bibfnamefont{G.}~\bibnamefont{Shiu}}
  (\bibinfo{year}{2015}), \eprint{1512.08984}.

\bibitem[{\citenamefont{Martin}(1997)}]{MSSMInt}
\bibinfo{author}{\bibfnamefont{S.~P.} \bibnamefont{Martin}}
  (\bibinfo{year}{1997}), \bibinfo{note}{[Adv. Ser. Direct. High Energy
  Phys.18,1(1998)]}, \eprint{hep-ph/9709356}.

\bibitem[{\citenamefont{Ellwanger et~al.}(2010)\citenamefont{Ellwanger,
  Hugonie, and Teixeira}}]{NMSSMInt}
\bibinfo{author}{\bibfnamefont{U.}~\bibnamefont{Ellwanger}},
  \bibinfo{author}{\bibfnamefont{C.}~\bibnamefont{Hugonie}}, \bibnamefont{and}
  \bibinfo{author}{\bibfnamefont{A.~M.} \bibnamefont{Teixeira}},
  \bibinfo{journal}{Phys. Rept.} \textbf{\bibinfo{volume}{496}},
  \bibinfo{pages}{1} (\bibinfo{year}{2010}), \eprint{0910.1785}.

\bibitem[{\citenamefont{Babu et~al.}(1991)\citenamefont{Babu, Pati, and
  Stremnitzer}}]{VL1}
\bibinfo{author}{\bibfnamefont{K.~S.} \bibnamefont{Babu}},
  \bibinfo{author}{\bibfnamefont{J.~C.} \bibnamefont{Pati}}, \bibnamefont{and}
  \bibinfo{author}{\bibfnamefont{H.}~\bibnamefont{Stremnitzer}},
  \bibinfo{journal}{Phys. Lett.} \textbf{\bibinfo{volume}{B256}},
  \bibinfo{pages}{206} (\bibinfo{year}{1991}).

\bibitem[{\citenamefont{Moroi and Okada}(1992{\natexlab{a}})}]{VL2}
\bibinfo{author}{\bibfnamefont{T.}~\bibnamefont{Moroi}} \bibnamefont{and}
  \bibinfo{author}{\bibfnamefont{Y.}~\bibnamefont{Okada}},
  \bibinfo{journal}{Mod. Phys. Lett.} \textbf{\bibinfo{volume}{A7}},
  \bibinfo{pages}{187} (\bibinfo{year}{1992}{\natexlab{a}}).

\bibitem[{\citenamefont{Moroi and Okada}(1992{\natexlab{b}})}]{VL3}
\bibinfo{author}{\bibfnamefont{T.}~\bibnamefont{Moroi}} \bibnamefont{and}
  \bibinfo{author}{\bibfnamefont{Y.}~\bibnamefont{Okada}},
  \bibinfo{journal}{Phys. Lett.} \textbf{\bibinfo{volume}{B295}},
  \bibinfo{pages}{73} (\bibinfo{year}{1992}{\natexlab{b}}).

\bibitem[{\citenamefont{Babu and Pati}(1996)}]{VL4}
\bibinfo{author}{\bibfnamefont{K.~S.} \bibnamefont{Babu}} \bibnamefont{and}
  \bibinfo{author}{\bibfnamefont{J.~C.} \bibnamefont{Pati}},
  \bibinfo{journal}{Phys. Lett.} \textbf{\bibinfo{volume}{B384}},
  \bibinfo{pages}{140} (\bibinfo{year}{1996}), \eprint{hep-ph/9606215}.

\bibitem[{\citenamefont{Bastero-Gil and Brahmachari}(2000)}]{VL5}
\bibinfo{author}{\bibfnamefont{M.}~\bibnamefont{Bastero-Gil}} \bibnamefont{and}
  \bibinfo{author}{\bibfnamefont{B.}~\bibnamefont{Brahmachari}},
  \bibinfo{journal}{Nucl. Phys.} \textbf{\bibinfo{volume}{B575}},
  \bibinfo{pages}{35} (\bibinfo{year}{2000}), \eprint{hep-ph/9907318}.

\bibitem[{\citenamefont{Shafi and Tavartkiladze}(2000)}]{VL6}
\bibinfo{author}{\bibfnamefont{Q.}~\bibnamefont{Shafi}} \bibnamefont{and}
  \bibinfo{author}{\bibfnamefont{Z.}~\bibnamefont{Tavartkiladze}},
  \bibinfo{journal}{Nucl. Phys.} \textbf{\bibinfo{volume}{B580}},
  \bibinfo{pages}{83} (\bibinfo{year}{2000}), \eprint{hep-ph/9909238}.

\bibitem[{\citenamefont{Babu et~al.}(2004)\citenamefont{Babu, Gogoladze, and
  Kolda}}]{VL7}
\bibinfo{author}{\bibfnamefont{K.~S.} \bibnamefont{Babu}},
  \bibinfo{author}{\bibfnamefont{I.}~\bibnamefont{Gogoladze}},
  \bibnamefont{and} \bibinfo{author}{\bibfnamefont{C.}~\bibnamefont{Kolda}}
  (\bibinfo{year}{2004}), \eprint{hep-ph/0410085}.

\bibitem[{\citenamefont{Barger et~al.}(2007)\citenamefont{Barger, Jiang,
  Langacker, and Li}}]{VL8}
\bibinfo{author}{\bibfnamefont{V.}~\bibnamefont{Barger}},
  \bibinfo{author}{\bibfnamefont{J.}~\bibnamefont{Jiang}},
  \bibinfo{author}{\bibfnamefont{P.}~\bibnamefont{Langacker}},
  \bibnamefont{and} \bibinfo{author}{\bibfnamefont{T.}~\bibnamefont{Li}},
  \bibinfo{journal}{Int. J. Mod. Phys.} \textbf{\bibinfo{volume}{A22}},
  \bibinfo{pages}{6203} (\bibinfo{year}{2007}), \eprint{hep-ph/0612206}.

\bibitem[{\citenamefont{Babu et~al.}(2008)\citenamefont{Babu, Gogoladze,
  Rehman, and Shafi}}]{VL9}
\bibinfo{author}{\bibfnamefont{K.~S.} \bibnamefont{Babu}},
  \bibinfo{author}{\bibfnamefont{I.}~\bibnamefont{Gogoladze}},
  \bibinfo{author}{\bibfnamefont{M.~U.} \bibnamefont{Rehman}},
  \bibnamefont{and} \bibinfo{author}{\bibfnamefont{Q.}~\bibnamefont{Shafi}},
  \bibinfo{journal}{Phys. Rev.} \textbf{\bibinfo{volume}{D78}},
  \bibinfo{pages}{055017} (\bibinfo{year}{2008}), \eprint{0807.3055}.

\bibitem[{\citenamefont{Graham et~al.}(2010)\citenamefont{Graham, Ismail,
  Rajendran, and Saraswat}}]{VL10}
\bibinfo{author}{\bibfnamefont{P.~W.} \bibnamefont{Graham}},
  \bibinfo{author}{\bibfnamefont{A.}~\bibnamefont{Ismail}},
  \bibinfo{author}{\bibfnamefont{S.}~\bibnamefont{Rajendran}},
  \bibnamefont{and} \bibinfo{author}{\bibfnamefont{P.}~\bibnamefont{Saraswat}},
  \bibinfo{journal}{Phys. Rev.} \textbf{\bibinfo{volume}{D81}},
  \bibinfo{pages}{055016} (\bibinfo{year}{2010}), \eprint{0910.3020}.

\bibitem[{\citenamefont{Martin}(2010{\natexlab{a}})}]{VL11}
\bibinfo{author}{\bibfnamefont{S.~P.} \bibnamefont{Martin}},
  \bibinfo{journal}{Phys. Rev.} \textbf{\bibinfo{volume}{D81}},
  \bibinfo{pages}{035004} (\bibinfo{year}{2010}{\natexlab{a}}),
  \eprint{0910.2732}.

\bibitem[{\citenamefont{Martin}(2010{\natexlab{b}})}]{VL12}
\bibinfo{author}{\bibfnamefont{S.~P.} \bibnamefont{Martin}},
  \bibinfo{journal}{Phys. Rev.} \textbf{\bibinfo{volume}{D82}},
  \bibinfo{pages}{055019} (\bibinfo{year}{2010}{\natexlab{b}}),
  \eprint{1006.4186}.

\bibitem[{\citenamefont{Liu}(2009)}]{VL13}
\bibinfo{author}{\bibfnamefont{C.}~\bibnamefont{Liu}}, \bibinfo{journal}{Phys.
  Rev.} \textbf{\bibinfo{volume}{D80}}, \bibinfo{pages}{035004}
  (\bibinfo{year}{2009}), \eprint{0907.3011}.

\bibitem[{\citenamefont{Liu and Lu}(2013)}]{VL14}
\bibinfo{author}{\bibfnamefont{C.}~\bibnamefont{Liu}} \bibnamefont{and}
  \bibinfo{author}{\bibfnamefont{J.-S.} \bibnamefont{Lu}},
  \bibinfo{journal}{JHEP} \textbf{\bibinfo{volume}{05}}, \bibinfo{pages}{040}
  (\bibinfo{year}{2013}), \eprint{1305.0070}.

\bibitem[{\citenamefont{Chang et~al.}(2013)\citenamefont{Chang, Liu, and
  Tang}}]{VL15}
\bibinfo{author}{\bibfnamefont{X.}~\bibnamefont{Chang}},
  \bibinfo{author}{\bibfnamefont{C.}~\bibnamefont{Liu}}, \bibnamefont{and}
  \bibinfo{author}{\bibfnamefont{Y.-L.} \bibnamefont{Tang}},
  \bibinfo{journal}{Phys. Rev.} \textbf{\bibinfo{volume}{D87}},
  \bibinfo{pages}{075012} (\bibinfo{year}{2013}), \eprint{1303.7055}.

\bibitem[{\citenamefont{Bonne and Moreau}(2012)}]{VLAdded1}
\bibinfo{author}{\bibfnamefont{N.}~\bibnamefont{Bonne}} \bibnamefont{and}
  \bibinfo{author}{\bibfnamefont{G.}~\bibnamefont{Moreau}},
  \bibinfo{journal}{Phys. Lett.} \textbf{\bibinfo{volume}{B717}},
  \bibinfo{pages}{409} (\bibinfo{year}{2012}), \eprint{1206.3360}.

\bibitem[{\citenamefont{Angelescu
  et~al.}(2015{\natexlab{b}})\citenamefont{Angelescu, Djouadi, and
  Moreau}}]{VLAdded2}
\bibinfo{author}{\bibfnamefont{A.}~\bibnamefont{Angelescu}},
  \bibinfo{author}{\bibfnamefont{A.}~\bibnamefont{Djouadi}}, \bibnamefont{and}
  \bibinfo{author}{\bibfnamefont{G.}~\bibnamefont{Moreau}}
  (\bibinfo{year}{2015}{\natexlab{b}}), \eprint{1510.07527}.

\bibitem[{\citenamefont{Moreau}(2013)}]{VLAdded3}
\bibinfo{author}{\bibfnamefont{G.}~\bibnamefont{Moreau}},
  \bibinfo{journal}{Phys. Rev.} \textbf{\bibinfo{volume}{D87}},
  \bibinfo{pages}{015027} (\bibinfo{year}{2013}), \eprint{1210.3977}.

\bibitem[{\citenamefont{Alves et~al.}(2015{\natexlab{b}})\citenamefont{Alves,
  Camargo, and Dias}}]{VLAdded4}
\bibinfo{author}{\bibfnamefont{A.}~\bibnamefont{Alves}},
  \bibinfo{author}{\bibfnamefont{D.~A.} \bibnamefont{Camargo}},
  \bibnamefont{and} \bibinfo{author}{\bibfnamefont{A.~G.} \bibnamefont{Dias}}
  (\bibinfo{year}{2015}{\natexlab{b}}), \eprint{1511.04449}.

\bibitem[{\citenamefont{Aad et~al.}(2012)}]{Higgs1}
\bibinfo{author}{\bibfnamefont{G.}~\bibnamefont{Aad}} \bibnamefont{et~al.}
  (\bibinfo{collaboration}{ATLAS}), \bibinfo{journal}{Phys. Lett.}
  \textbf{\bibinfo{volume}{B716}}, \bibinfo{pages}{1} (\bibinfo{year}{2012}),
  \eprint{1207.7214}.

\bibitem[{\citenamefont{Chatrchyan et~al.}(2012)}]{Higgs2}
\bibinfo{author}{\bibfnamefont{S.}~\bibnamefont{Chatrchyan}}
  \bibnamefont{et~al.} (\bibinfo{collaboration}{CMS}), \bibinfo{journal}{Phys.
  Lett.} \textbf{\bibinfo{volume}{B716}}, \bibinfo{pages}{30}
  (\bibinfo{year}{2012}), \eprint{1207.7235}.

\bibitem[{\citenamefont{Djouadi}(2008{\natexlab{a}})}]{HEFT1}
\bibinfo{author}{\bibfnamefont{A.}~\bibnamefont{Djouadi}},
  \bibinfo{journal}{Phys. Rept.} \textbf{\bibinfo{volume}{457}},
  \bibinfo{pages}{1} (\bibinfo{year}{2008}{\natexlab{a}}),
  \eprint{hep-ph/0503172}.

\bibitem[{\citenamefont{Djouadi}(2008{\natexlab{b}})}]{HEFT2}
\bibinfo{author}{\bibfnamefont{A.}~\bibnamefont{Djouadi}},
  \bibinfo{journal}{Phys. Rept.} \textbf{\bibinfo{volume}{459}},
  \bibinfo{pages}{1} (\bibinfo{year}{2008}{\natexlab{b}}),
  \eprint{hep-ph/0503173}.

\bibitem[{\citenamefont{Giudice and Rattazzi}(1999)}]{GMSB_GUT}
\bibinfo{author}{\bibfnamefont{G.~F.} \bibnamefont{Giudice}} \bibnamefont{and}
  \bibinfo{author}{\bibfnamefont{R.}~\bibnamefont{Rattazzi}},
  \bibinfo{journal}{Phys. Rept.} \textbf{\bibinfo{volume}{322}},
  \bibinfo{pages}{419} (\bibinfo{year}{1999}), \eprint{hep-ph/9801271, see page
  9}.

\bibitem[{\citenamefont{Olive et~al.}(2014)}]{PDG}
\bibinfo{author}{\bibfnamefont{K.~A.} \bibnamefont{Olive}} \bibnamefont{et~al.}
  (\bibinfo{collaboration}{Particle Data Group}), \bibinfo{journal}{Chin.
  Phys.} \textbf{\bibinfo{volume}{C38}}, \bibinfo{pages}{090001}
  (\bibinfo{year}{2014}).

\bibitem[{\citenamefont{Ellwanger et~al.}(2005)\citenamefont{Ellwanger, Gunion,
  and Hugonie}}]{NMSSMTools1}
\bibinfo{author}{\bibfnamefont{U.}~\bibnamefont{Ellwanger}},
  \bibinfo{author}{\bibfnamefont{J.~F.} \bibnamefont{Gunion}},
  \bibnamefont{and} \bibinfo{author}{\bibfnamefont{C.}~\bibnamefont{Hugonie}},
  \bibinfo{journal}{JHEP} \textbf{\bibinfo{volume}{02}}, \bibinfo{pages}{066}
  (\bibinfo{year}{2005}), \eprint{hep-ph/0406215}.

\bibitem[{\citenamefont{Ellwanger and Hugonie}(2006)}]{NMSSMTools2}
\bibinfo{author}{\bibfnamefont{U.}~\bibnamefont{Ellwanger}} \bibnamefont{and}
  \bibinfo{author}{\bibfnamefont{C.}~\bibnamefont{Hugonie}},
  \bibinfo{journal}{Comput. Phys. Commun.} \textbf{\bibinfo{volume}{175}},
  \bibinfo{pages}{290} (\bibinfo{year}{2006}), \eprint{hep-ph/0508022}.

\bibitem[{\citenamefont{Belanger et~al.}(2005)\citenamefont{Belanger, Boudjema,
  Hugonie, Pukhov, and Semenov}}]{NMSSMTools3}
\bibinfo{author}{\bibfnamefont{G.}~\bibnamefont{Belanger}},
  \bibinfo{author}{\bibfnamefont{F.}~\bibnamefont{Boudjema}},
  \bibinfo{author}{\bibfnamefont{C.}~\bibnamefont{Hugonie}},
  \bibinfo{author}{\bibfnamefont{A.}~\bibnamefont{Pukhov}}, \bibnamefont{and}
  \bibinfo{author}{\bibfnamefont{A.}~\bibnamefont{Semenov}},
  \bibinfo{journal}{JCAP} \textbf{\bibinfo{volume}{0509}}, \bibinfo{pages}{001}
  (\bibinfo{year}{2005}), \eprint{hep-ph/0505142}.

\end{thebibliography}

\end{document}